\documentclass{mn2e}
\usepackage{graphicx}

\def\go{
\mathrel{\raise.3ex\hbox{$>$}\mkern-14mu\lower0.6ex\hbox{$\sim$}}
}
\def\lo{
\mathrel{\raise.3ex\hbox{$<$}\mkern-14mu\lower0.6ex\hbox{$\sim$}}
}
\def\simeq{
\mathrel{\raise.3ex\hbox{$\sim$}\mkern-14mu\lower0.4ex\hbox{$-$}}
}

\def\etal{{et al.\ }}

\begin{document}

\title[{\it XMM-Newton} spectroscopy of high-redshift quasars]
{{\it XMM-Newton} spectroscopy of high-redshift quasars}
\author[K.L. Page \etal]{K.L. Page$^{1}$, J.N. Reeves$^{2,3}$, P.T. O'Brien$^{1}$ and M.J.L. Turner$^{1}$\\
$^{1}$ X-Ray and Observational Astronomy Group, Department of Physics \& Astronomy, University of Leicester, LE1 7RH, UK\\
$^{2}$ Laboratory for High Energy Astrophysics, Code 662, NASA Goddard Space Flight Center, Greenbelt, MD 20771, USA\\
$^{3}$ Department of Physics \& Astronomy, Johns Hopkins University, 3400 N. Charles Street, Baltimore, MD 21218, USA\\
}

\label{firstpage}

\maketitle

\begin{abstract}

{\it XMM-Newton} observations of 29 high redshift (z$>$2) quasars, including seven radio-quiet, 16 radio-loud and six Broad Absorption Line (BAL) objects, are presented; due to the high redshifts, the rest-frame energy bands extend up to $\sim$~30--70~keV. Over 2--10~keV, the quasars can be well fitted in each case by a simple power-law, with no strong evidence for iron emission lines. The lack of iron lines is in agreement both with dilution by the radio jet emission (for the radio-loud quasars) and the X-ray Baldwin effect. No Compton reflection humps at higher energies (i.e., above 10~keV in the rest frame) are detected either.
Over the broad-band (0.3--10~keV), approximately half (nine out of 16) of the radio-loud quasars are intrinsically absorbed, with the values of N$_{\rm H}$ generally being 1--2~$\times$~10$^{22}$ cm$^{-2}$ in the rest frames of the objects. None of the seven radio-quiet objects shows excess absorption, while four of the six BAL quasars are absorbed.
The radio-loud quasars have flatter continuum slopes than their radio-quiet counterparts ($\Gamma_{\rm RL}$~$\sim$~1.55; $\Gamma_{\rm RQ}$~$\sim$~1.98 over 2--10~keV), while, after modelling the absorption, the underlying photon index for the six BAL quasars is formally consistent with the non-BAL radio-quiet objects.

\end{abstract}


\date{Received / Accepted}


\section{Introduction}
\label{intro}

Approximately 10~per~cent of all active galactic nuclei (AGN) are classified as being radio-loud; that is, the logarithm of the ratio of the radio (5~GHz) to optical ($B$-band, 4400\AA) fluxes, R$_{\rm L}$, is $\geq$~1 (Wilkes \& Elvis 1987; Kellerman \etal 1989). In general, for a given optical luminosity, the X-ray emission from radio-loud quasars (RLQs) is about three times greater than that from their radio-quiet counterparts (RQQs; Zamorani \etal 1981; Worrall \etal 1987), which allows them to be studied out to higher redshifts. The high effective area of the EPIC (European Photon Imaging Camera; Turner \etal 2001; Str{\" u}der \etal 2001) instruments on-board {\it XMM-Newton} is enabling the detection of an ever-increasing number of high-z Quasi-Stellar Objects (QSOs), both radio-quiet and radio-loud.

In Page \etal (2004a), the authors investigated the X-ray spectra of a sample of high-luminosity RQQs, finding that the continuum shape is essentially the same as for lower-luminosity (and, presumably, lower black-hole mass) radio-quiet AGN, i.e. Seyfert galaxies. The broad-band X-ray spectra could be well modelled either by a power-law ($\Gamma$~$\sim$~1.9) together with one or two blackbody components to parametrize the soft excess, or by a double Comptonisation model (one component modelling the high-energy power-law; the other, the soft excess). Some of the RQQs showed evidence for neutral iron emission, although it has previously been found that the strength of the narrow Fe K$\alpha$ line diminishes as the X-ray luminosity increases (Iwasawa \& Taniguchi 1993; Nandra \etal 1997b; Page \etal 2004b); this is the so-called X-ray Baldwin effect. In RLQs, these spectral features may be somewhat diluted, since some of the harder X-ray emission is thought to be Synchrotron or inverse Compton emission from the relativistic radio jet, rather than from the accretion disc. This also leads to RLQs tending to have flatter spectra over the 2--10~keV rest-frame (e.g., Williams \etal 1992; Lawson \etal 1997; Reeves \etal 1997; Reeves \& Turner 2000). 

This paper aims to analyse the X-ray spectra of a sample of high-redshift QSOs, both radio-quiet and radio-loud, investigating, in particular, the continuum shape, iron emission, Compton reflection and absorption. The results of such a comparison should help to determine just how fundamentally different RLQs and RQQs are: are the central engines, themselves, different, or are the differences solely due to the radio jet?  A further major aim of the paper is to investigate how the X-ray spectral and continuum properties of high-redshift quasars (z~=~2--6) compares to those at lower redshifts. For instance, is there any evolution of the X-ray spectra with either luminosity or redshift, which might suggest that properties of quasar central engines evolve with time? Does the continuum photon index, or amount of intrinsic absorption, vary, suggesting a changing environment?  

Section~\ref{obs} explains how the data were processed and the spectra obtained, while Section~\ref{specan} details the spectral analysis. The results are discussed in Section~\ref{disc}.


\section{XMM-Newton observations}
\label{obs}

The sample in this paper constitutes all available {\it XMM-Newton} observations (until the end of 2004) of z$>$2 QSOs, with sufficient counts to allow spectral fitting (approximately 500 counts in the combined MOS+PN results). 

\begin{figure}
\begin{center}
\includegraphics[clip,width=6.4cm,angle=-90]{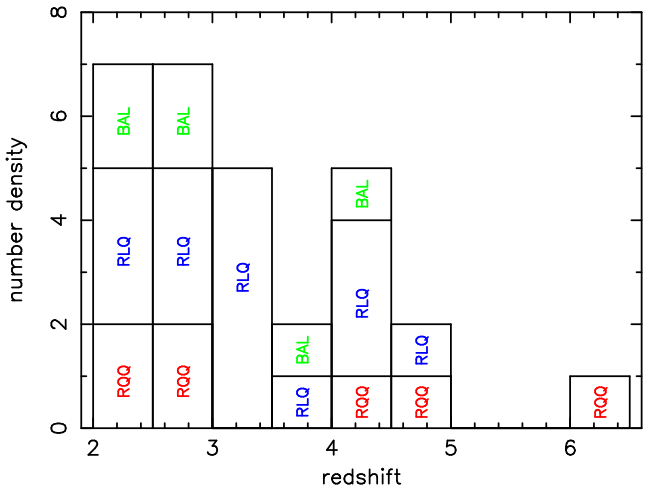}
\caption{A histogram showing the distribution of the redshifts for the RLQ, RQQs and BAL QSOs.}
\label{histogram}
\end{center}
\end{figure}

Table~\ref{xmm} lists the QSOs, their right ascensions and declinations, redshifts, Galactic absorbing columns (determined from the {\sc ftool} {\bf nh}; Dickey \& Lockman 1990) and radio data, as well as information about the actual {\it XMM-Newton} observations. Where the NRAO/VLA Sky Survey (NVSS) did not detect the object at 1.4~GHz, the survey limit of 2.5~mJy is given, except for four cases (Q~0000$-$263, SDSS~1030+0524, SBS~1107+487 and APM~08279+5255) where more stringent limits were found in the literature. Figure~\ref{histogram} plots the redshift distribution for the objects.
The QSOs are a combination of proprietary data and public observations obtained from the {\it XMM-Newton} Science Archive\footnote{http://xmm.vilspa.esa.es/external/xmm\_data\_acc/xsa/index.shtml}.

\begin{table*}
\begin{center}
\caption{Information about the QSOs and the {\it XMM-Newton} observations; the radio-loud and radio-quiet objects are separately ordered by Right Ascension. $^{a}$ Galactic N$_{\rm H}$, determined from the {\sc ftool} {\bf nh}; $^{b}$ from NVSS - NRAO/VLA Sky Survey (Condon \etal 1998). PKS 0438$-$43 is outside the region covered; $^{c}$ R$_{\rm L}$~=~F$_{5GHz}$/F$_B$; $^{d}$ Vignali, Brandt \& Schneider (2003); $^{e}$ Kuhn \etal (2001); $^{f}$ Turnshek \etal (1991);  $^{g}$ Holt \etal (2004); (1) -- values from Reeves \& Turner (2000); (2) -- 5~GHz from Gregory \& Condon (1991); the B-band flux was calculated from NED; (3) -- Elvis \etal (1994); (4) -- Kaspi, Brandt \& Schneider (2000); (5) -- 5~GHz from Gregory \etal (1996); the B-band flux was calculated from NED; (6) -- calculated from Zickgraf \etal (1997) and the NASA/IPAC Extragalactic Database (NED); (7) -- for these objects, the 5~GHz upper limit was estimated from the 1.4~GHz value, using f$_{\nu}$~$\propto$~$\nu^{-0.8}$ for the radio power-law; the B-band flux was calculated from NED; (8) -- calculated from Lonsdale, Barthel \& Miley (1993) and NED; (9) -- calculated from Parkes-MIT-NRAO (PMN) survey (Wright \etal 1996) and Fabian \etal (2001); (10) --  5~GHz upper limit estimated from the 1.4~GHz value; B-band flux from Bechtold \etal (2003); (11) -- Vignali, Brandt \& Schneider (2003); (12) -- Kuhn \etal (2001); (13)  --  5~GHz upper limit estimated from the 1.4~GHz value; B-band flux from Dai \etal (2004); (14) -- 5~GHz upper limit estimated from the 1.4~GHz value; B-band flux from Grupe, Mathur \& Elvis (2003)}

\label{xmm}
\begin{tabular}{p{2.4truecm}p{1.1truecm}p{1.4truecm}p{1.4truecm}p{1.4truecm}p{0.9truecm}p{0.9truecm}p{0.5truecm}p{0.6truecm}p{1.05truecm}p{1.25truecm}p{1.6truecm}}
\hline
Object & RA & Dec & Obs. ID & Obs. date & \multicolumn{3}{c} {exposure time (ks)} & z & Gal. N$_{\rm H}^{a}$ & 1.4~GHz  & log(R$_{\rm L})^{c}$ \\
 &   (J2000)& (J2000) & &  & MOS 1 & MOS 2 & PN  & & (10$^{20}$  & radio  &    \\
& & & & & &  & && cm$^{-2}$)& flux$^{b}$  (mJy) & \\
\hline
{\bf RADIO LOUD}\\

S5 0014+81 & 00:17:08.0 & +81:35:08.0 & 0112620201 & 2001-08-23 & 38.9 & 39.0 & 32.8  & 3.366 & 13.9 & 693~$\pm$~21 & 2.33 (1) \\
RBS 315 & 02:25:04.7 & +18:46:49.0 & 0150180101 & 2003-07-25 & 21.6 & 21.7 & 18.3 & 2.69 & 10.0 & 461~$\pm$~14 & 3.47 (2)\\

Q~0420$-$388 & 04:22:14.8 & $-$38:44:53.0 & 0104860101 & 2003-01-14 & 10.2 & 10.7 & 8.1 & 3.11 & 2.1 & 197~$\pm$~6 & 2.24 (3)\\
PKS 0438$-$43 & 04:40:17.2 & $-$43:33:08.6 & 0104860201 & 2002-04-06 &12.2 &12.2 & 9.0&2.863 & 1.8 & --- & 4.75 (1) \\
PMN 0525$-$343 & 05:25:06.2 & $-$33:43:05.5 & 0050150301 &2001-09-15 & 24.2 & 24.2&19.8 & 4.413 & 2.2 & 188~$\pm$~6 & 3.20 (4)\\
PKS~0537-286 & 05:39:54.2 & $-$28:39:55.0 & 0114090101 & 2000-03-19 & 19.2 & 28.0 & 40.0 & 3.104 & 2.1 & 862~$\pm$~26 & 4.35 (1)\\
S5 0836+71 & 08:41:24.0 & +70:53:41.0 & 0112620101 & 2001-04-13 & 25.8 & 25.5 & 22.9  & 2.172 & 2.9 & 9329~$\pm$~115 & 3.33 (1)\\
HS 0848+1119 & 08:50:45.7 & +11:08:40.0 & 0148742501 & 2003-10-28 & 8.3 & 8.4 & 5.5 & 2.62 & 4.3 & 50~$\pm$~2 & 1.77 (5)\\
RX J1028.6$-$0844 & 10:28:38.7 & $-$08:44:38.8 & 0153290101 &2003-06-13 & 26.2 &27.0 & 17.7 & 4.276 & 4.6 & 269~$\pm$~8 & 3.61 (6)\\
RX J122135.6 & 12:21:35.6 & +28:06:14.0 & 0104860501 & 2002-06-26 & 28.9 & 30.3 & 26.6& 3.305 & 1.9 & 26.4~$\pm$~0.9 & 2.34 (7)\\
+280613 \\
SBS~1345+584 & 13:47:40.9 & +58:12:43.0 & 0112250201 & 2001-06-06 & 32.7 & 32.7  & 25.0 & 2.039 & 1.3 & 642~$\pm$~23 & 2.41 (8)\\
87GB 142825.9+421804  & 14:30:23.7 & +42:04:36.5 & 0111260701 & 2003-01-17 & 14.2 & 14.2 & 11.5 & 4.715 & 1.4 & 211~$\pm$~6 & 3.28 (9)\\
87GB 150844.6+571424 & 15:10:02.9 & +57:02:43.4 & 0111260201 & 2002-05-11&11.4 &11.5 &8.4 &4.309 &1.5 & 202~$\pm$~6 & 3.36 (1) \\
PKS 2000$-$330 & 20:03:24.1 & $-$32:51:45.1 & 0104860601 & 2002-04-14&15.3 &15.3 &10.5 &3.773 &7.9 & 446~$\pm$~16 &4.05 (1)\\
PKS 2126$-$15 & 21:29:12.2 & $-$15:38:41.0 & 0103060101 & 2001-05-01 & 19.1 & 19.2 & 13.4 & 3.268 & 5.0 & 590~$\pm$~18 & 3.37 (1) \\
PKS 2149$-$306 & 21:51:55.6 & $-$30:27:53.7 & 0103060401 & 2001-05-01 & 19.1 & 19.1 & 11.5 & 2.345 & 2.1 & 1249~$\pm$~37 & 3.83 (1)\\

{\bf RADIO QUIET}\\

LBQS~0053$-$0134 & 00:56:14.5 & $-$01:18:28.0 & 0012440101 & 2001-01-14 & 28.9 & 28.9 & 24.0 & 2.062 & 3.2 & $<$2.5 & $<$$-$0.68 (7)\\

PSS~J0248+1802 & 02:48:54.3 & +18:02:50.0 & 0150870301 & 2003-02-14 & 8.5 & 9.3 & 9.0 & 4.411 & 9.2 & $<$2.5 & $<$0.54 (10)\\

SDSS~1030+0524 & 10:30:27.1 & +05:24:55.0 & 0148560501 & 2003-05-22 & 71.1 &72.4 &56.3 & 6.28 & 3.2 & $<$0.46$^{d}$ & $<$0.97 (11)\\

HE~1104$-$1805 & 11:06:33.0 & $-$18:21:24.0 & 0112630101 & 2001-06-14 & 12.1 & 12.4 & 8.6 & 2.305 & 4.6 &  $<$2.5 & $<$$-$0.96 (1) \\

SBS~1107+487 & 11:10:38.5 & +48:31:16.0 & 0059750401 & 2002-04-25 & 21.2 & 21.3  & 14.1  & 2.958 & 1.8 & $\sim$0.33$^{e}$ & $<$$-$0.68 (12) \\


Q~1442+2931 & 14:44:53.7 & +29:19:05.2 & 0103060201 & 2002-08-01& 21.8 & 21.8 & 14.6 & 2.638 & 1.6 &$<$2.5 & $<$$-$0.19 (7)\\

BR~2237$-$0607 & 22:39:53.6 & $-$05:52:20.0 & 0149410401 & 2003-05-17 & 25.8 & 26.8 & 19.1 & 4.558 & 3.8 &$<$2.5 & $<$$-$0.73 (7)\\

{\bf BAL}\\

Q~0000$-$263 & 00:03:22.9 & $-$26:03:16.8 & 0103060301 & 2002-06-26 & 36.6 & 36.5 & 30.2 & 4.098 & 1.7 & $<$0.5$^{f}$ & $<$$-$0.10 (7) \\

APM 08279+5255 & 08:31:41.6 & +52:45:17.0 & 0092800101 & 2001-10-30 & 16.7 & 16.7 & 13.0 & 3.87 & 3.8 & 1.3$^{g}$ &  $<$0.23 (13)\\

RX~J0911.4+0551 & 09:11:27.5 & +05:50:52.0 & 0083240201 & 2001-11-02 & 17.7 & 17.0 & 12.6 & 2.80 & 3.6 & $<$2.5 & $<$0.07 (13)\\

Q 1246-057 & 12:49:13.8 &  $-$05:59:18.0 & 0060370201 & 2001-07-11 & 39.9 & 40.0 & 34.3 & 2.236 & 2.1 & $<$2.5 & $<$0.89 (14)\\

CSO 0755 & 15:25:53.9 & +51:36:49.4 & 0011830201 & 2001-12-08 &29.4 & 29.2 & 24.2& 2.86 & 1.6 & $<$2.5& $<$$-$0.19 (7)\\

SBS 1542+541 & 15:43:59.4 &  +53:59:03.0 & 0060370901 & 2002-02-06 & 20.9 & 20.8 & 15.2 & 2.371 & 1.3 & $<$2.5 &  $<$0.39 (14)\\

\hline
\end{tabular}
\end{center}
\end{table*}

The ODFs (Observation Data Files) were obtained from the XMM Science Archive and processed using {\sc sas} v5.4\footnote{The data for this paper were analysed during 2004, using the most recent calibration files available at the time.} for consistency, with the exception of PKS~0537$-$286. Because PKS~0537$-$286 was observed very early on in the mission (revolution 51), the ODFs are in a format which cannot be easily reprocessed. Hence, the results presented here are from the original data products; such an analysis was first carried out by Reeves \etal (2001). Before extracting the spectra, background light-curves were produced to identify any periods of high background. Most of the observations were found to show intermittent background flaring, so these time slots were subsequently ignored. However, the vast majority of the observation of S5 0014+81 was affected by high background, so excluding all the flares was deemed to be impractical. For this object, a very small region of 20~arcsec was used to extract the spectrum and, hence, minimise the background contribution. Similarly small regions were used for PMN~0525$-$3343, PKS~2000$-$330 and RX~J1028.6$-$0844 in addition to excluding the worst periods of high background.

For the other QSOs, spectra were then extracted from circular regions (40--50~arcsec in radius), with the background spectra obtained from offset `source-free' areas. Patterns 0--12 for MOS (0--4 for PN) were chosen, since there was no indication of pile-up. The {\sc sas} tasks {\sc rmfgen} and {\sc arfgen} were then run, to produce the redistribution matrix functions and the ancillary response functions respectively. MOS 1 and MOS 2 spectra were co-added, after determining there were no large discrepancies between them, and this combined spectrum was then fitted simultaneously with the PN data, using {\sc Xspec} version 11.3,  to produce the results given in this paper.

Throughout this paper, a WMAP (Wilkinson Microwave Anisotropy Probe) Cosmology of H$_0$~=~70~km~s$^{-1}$~Mpc$^{-1}$, $\Omega_{\lambda}$~=~0.73 and $\Omega_{m}$~=~1-$\Omega_{\lambda}$ is assumed. Errors are given at the 90~per~cent level (e.g., $\Delta\chi^{2}$~=~2.7 for one degree of freedom).

\section{Spectral Analysis}
\label{specan}
\subsection{Simple power-law fits}

As is conventional, each of the spectra was first fitted with a simple power law, over the 2--10 keV (rest-frame) band; the resulting photon indices are given in Table~\ref{hardfits}. None of the QSOs showed statistically significant iron emission (the highest F-test probability was $\sim$95~per~cent), so only the upper limits for the equivalent widths of narrow lines are given in the table. For many of the sample including such a Gaussian led to unconstrained line energies; in these cases, the energy was fixed at 6.4~keV, corresponding to neutral iron. 

The highest redshift (z~=~6.28) object, SDSS~1030+0524, showed a small amount of excess flux (one data bin in both the PN and MOS spectra) at approximately the correct energy for ionised iron emission, as previously noted by Farrah \etal (2004); Table~\ref{hardfits} states the results for fitting {\it neutral} narrow lines only. Fitting the possible feature with a Gaussian gave an energy of 7.02$^{+0.16}_{-0.37}$ keV; this was approximately 92~per~cent significant, with a high upper limit for the equivalent width of 3~keV.


\begin{table*}
\begin{center}
\caption{Fits to the rest-frame 2--10~keV band. The line width was fixed at $\sigma$~=~10~eV and, if the energy of the fitted line was unconstrained, it was fixed ($^{f}$) at 6.4~keV; the EW is given for the rest-frame of each object. The F-test null probability is stated for the inclusion of the line;
if there is no change in $\chi^{2}$, then the F-test is undefined.
Certain of the spectra were better fitted by the inclusion of excess (rest-frame) absorption; where this was the case, the value is given in the fifth column. $^{a}$ The spectrum of APM~08279+5255 also contains an ionised Fe~K edge at 7.70$^{+0.26}_{-0.31}$~keV, with an optical depth, $\tau$~=~0.44$^{+0.22}_{-0.21}$. This object is also lensed, with a magnification factor of $\sim$~142, as 
are RX~J0911+0551 (factor of 17) and HE~1104$-$1805 (factor of 12) -- values from Dai \etal (2004); the luminosities given in this table (marked as $^{**}$) are, however, {\it uncorrected}. The final column gives the limits on the reflection parameter: R~=~$\Omega$/2$\pi$, where $\Omega$ is the solid angle subtended by the scattering medium. These were determined from fits to the full 0.3--10~keV {\it XMM-Newton} band.} 
\label{hardfits}
\begin{tabular}{p{3.0truecm}p{2.0truecm}p{1.8truecm}p{1.2truecm}p{1.7truecm}p{1.0truecm}p{1.5truecm}p{1.85truecm}p{1.7truecm}}
\hline
QSO &  $\Gamma$ & Line Energy &  EW (eV)& N$_{H}$ & $\chi^{2}$/$\nu$ & F-test null & Lum. & R\\
&  & (keV) & & (10$^{22}$ cm$^{-2}$)& & prob. (dof) &  (10$^{46}$ erg s$^{-1}$)\\
\hline
{\bf RADIO LOUD}\\

S5~0014+81  & 1.39~$\pm$~0.03 & 6.4$^{f}$ & $<$18 & & 601/483 & 0.37 (1) &  17.5 &  $<$0.23\\
RBS~315& 1.22~$\pm$~0.03 & 6.4$^{f}$ & $<$4 & 1.74$^{+0.18}_{-0.24}$  & 625/576 & --- & 36.96 & $<$0.01\\
Q~0420$-$388 &  1.96~$\pm$~0.10 & 6.4$^{f}$ & $<$159 & &90/87 & 0.33 (1) & 2.9& $<$0.24\\
PKS~0438$-$43 &  1.57~$\pm$~0.04 & 6.4$^{f}$ & $<$33 & &319/321 & --- &  9.4& 0.62~$\pm$~0.28\\
PMN~0525$-$3343 &  1.43~$\pm$~0.05 & 6.4$^{f}$ & $<$47 & &259/252 & --- &  9.3&$<$0.01\\
PKS~0537$-$286 &  1.52~$\pm$~0.03 & 6.18~$\pm$~0.19 & $<$69 & &503/439 & 0.11 (2)  & 8.6 & $<$0.05\\
S5~0836+71 &  1.36~$\pm$~0.01 & 6.35~$^{+0.13}_{-0.20}$ & $<$22 & &752/669 & 0.05 (2)& 65.7 & $<$0.01\\
HS~0848+1119 & 2.16$^{+0.67}_{-0.55}$ & 6.4$^{f}$ & $<$1310 & & 7/10 & 0.26 (1) & 0.3& $<$2.9\\

RX J1028.6$-$0844  & 1.27~$\pm$~0.06 & 6.4$^{f}$ & $<$112 & &221/201 & 0.18 (1) & 6.7 & $<$0.42\\
RX J122135.6+280613 &  1.32~$\pm$~0.09 & 6.4$^{f}$ & $<$151 & &90/119 & 0.25 (1)& 1.3& $<$1.09 \\

SBS~1345+584 &  2.07~$\pm$~0.10 & 6.35$^{+0.14}_{-0.17}$ & $<$289 & &115/104 & 0.11 (2)& 0.5& $<$0.71\\ 
87GB 142825.9+421804   &1.51~$\pm$~0.05 & 6.4$^{f}$ & $<$101 & &224/242 & 0.14 (1)& 15.2 & $<$0.15\\
87GB 150844.6+571424 &  1.60~$\pm$~0.13 & 6.4$^{f}$ & $<$117 & &65/58 & --- & 3.1 & $<$0.81\\
PKS~2000$-$330 &  1.67~$\pm$~0.10 & 6.4$^{f}$ & $<$55 & &102/107 & --- &  4.8& $<$0.37\\
PKS~2126$-$15  & 1.28~$\pm$~0.02 & 6.4$^{f}$ & $<$~21 & &663/496 & ---&  38.1& $<$0.01\\
PKS~2149$-$306 &  1.51~$\pm$~0.03 & 6.4$^{f}$ & $<$~32 & &536/508 & --- &  12.4& $<$0.05\\

{\bf RADIO QUIET}\\

LBQS~0053$-$0134 &  1.91$^{+0.82}_{-0.36}$ & 6.4$^{f}$ & $<$769 & &18/20 & --- & 0.07& $<$5.60\\
PSS~J0248+1802 & 2.23$^{+0.21}_{-0.20}$ & 6.4$^{f}$ & $<$260 & & 26/26 & --- & 1.5& $<$0.35\\

SDSS~1030+0524 &  2.12~$\pm$~0.47 & 6.4$^{f}$ & $<$874 & &14/7 & --- & 0.38& unconstrained\\
HE 1104$-$1805 &  1.90~$\pm$~0.25 & 6.4$^{f}$ & $<$301 & &23/25 & --- & 0.4$^{**}$ & $<$3.54 \\
SBS~1107+487 &  1.79$^{+0.52}_{-0.47}$ & 6.4$^{f}$ & $<$800 & &28/28 & --- & 0.4 & $<$4.62\\


Q~1442+2931 &   1.95~$\pm$~0.12 & 6.4$^{f}$ & $<$164 & &82/80 & --- & 0.8 & $<$1.06\\

BR~2237$-$0607  & 1.97$^{+0.37}_{-0.28}$ & 6.4$^{f}$ & $<$281 & & 14/12 & --- & 0.5& $<$2.56\\

{\bf BAL}\\

Q~0000$-$263 &   2.09~$\pm$~0.16 & 6.4$^{f}$ & $<$178 & & 38/55 & --- & 0.8& $<$2.82\\

APM 08279+5255$^{a}$ & 1.91$^{+0.30}_{-0.29}$ & 6.4$^{f}$ & $<$127 & 5.80$^{+1.77}_{-1.64}$ & 145/129 & 0.19 (2) & 6.15$^{**}$ & 3.24$^{+1.81}_{-3.23}$\\

RX~J0911.4+0551 & 1.12$^{+0.24}_{-0.23}$ & 6.51$^{+0.17}_{-0.10}$ & $<$585 & & 41/37 & 0.12 (2)& 0.5$^{**}$& $<$0.63\\

Q 1246-057 & 1.53$^{+0.35}_{-0.34}$ & 6.4$^{f}$ & $<$405 & & 28/37 & --- & 0.08& $<$2.3\\

CSO 0755 &  1.48~$\pm$~0.09 & 6.4$^{f}$ & $<$82 & &72/97 & --- & 0.9& $<$0.39\\

SBS 1542+541 & 2.2$^{+0.44}_{-0.41}$ & 6.4$^{f}$ & $<$202 & 3.29$^{+2.07}_{-1.94}$ & 39/36 & --- & 0.4& $<$0.14\\
\hline
\end{tabular}
\end{center}
\end{table*}

\begin{table*}
\begin{center}

\caption{Broad-band (0.3--10~keV, frame of observer) fits. The values given for N$_{\rm H}$ are in the rest-frames of the QSOs and are in additional to the Galactic values. $\Delta\chi^{2}$ gives the decrease in $\chi^{2}$ for the addition of the excess absorption. The final column gives the rest-frame energy bands corresponding to 0.3--10~keV in the observer's frame. $^{a}$ The fit to APM 08279+5255 also includes an ionised absorption edge as detailed for the 2--10~keV fit. $^{**}$ Again, the luminosities given in this table for  APM 08279+5255, RX~J0911+0551 and HE~1104$-$1805 are not corrected for the magnification by gravitational lensing (Dai \etal 2004).} 
\label{fits}
\begin{tabular}{p{3.0truecm}p{1.4truecm}p{1.7truecm}p{1.5truecm}p{1.2truecm}p{2.0truecm}p{1.6truecm}p{1.9truecm}p{1.7truecm}}
\hline
QSO & $\Gamma$ & N$_{\rm H}$ & $\chi^{2}$/$\nu$ & $\Delta\chi^{2}$ & F-test null & Lum. (10$^{46}$ & Flux (10$^{-12}$ &  Rest-frame \\
&  & (10$^{22}$ cm$^{-2}$) & & & prob. (dof)  &  erg s$^{-1}$) & erg cm$^{-2}$ s$^{-1}$) & band (keV)\\
\hline
{\bf RADIO LOUD}\\
S5~0014+81  & 1.61~$\pm$~0.02 & 1.82~$\pm$~0.19 & 1261/1159 & 345 & 1.0~$\times$~10$^{-99}$ (1) &59.4 & 5.59 & 1.3--44\\

RBS~315& 1.24~$\pm$~0.01 &  1.78$^{+0.08}_{-0.09}$ & 1819/1737 & 1867 & 1.0~$\times$~10$^{-99}$ (1)& 126.4 & 18.65& 1.1--37\\

Q~0420$-$388 & 1.97~$\pm$~0.08 & $<$0.76 & 129/125 & 6 & 1.7~$\times$~10$^{-2}$ (1) & 6.5 & 0.645 & 1.2--41\\

PKS~0438$-$43 & 1.73~$\pm$~0.03 & 1.18~$\pm$~0.14 & 482/507 & 246 & 1.0~$\times$~10$^{-99}$ (1) & 25.5 & 2.23& 1.2--39\\

PMN~0525$-$3343 & 1.73~$\pm$~0.04 & 1.91~$\pm$~0.38 & 466/433 & 86 & 1.1~$\times$~10$^{-17}$ (1) & 35.1& 2.93& 1.6--54\\

PKS~0537$-$286 & 1.41~$\pm$~0.01 & $<$0.04 & 1048/849 & 0 & --- & 30.1 & 2.93& 1.2--41\\

S5~0836+71 & 1.37~$\pm$~0.01 & 0.10~$\pm$~0.02 & 2585/2221 & 131 & 1.1~$\times$~10$^{-25}$ (1) & 195.2 & 21.52 & 1.0--32\\

HS~0848+1119 & 2.54~$^{+0.87}_{-0.61}$ & $<$4.70 & 14/18 & 2 & 0.126 & 0.6 & 0.08& 1.1--36\\

RX J1028.6$-$0844 & 1.46~$\pm$~0.04 & 1.28~$\pm$~0.44 & 371/362 & 26 & 7.5~$\times$~10$^{-7}$ & 26.2 & 1.25& 1.6--53\\

RX J122135.6+280613  & 1.38~$\pm$~0.06 & 0.64~$\pm$~0.37 & 173/196 & 9 & 1.6~$\times$~10$^{-3}$ (1) & 4.8 & 0.440&  1.3--43 \\

SBS~1345+584  & 2.00~$\pm$~0.07 &  $<$0.21 & 196/170 & 1 & 0.35 (1)& 1.0 & 0.315&  0.9--30\\

87GB 142825.9+421804 &  1.75~$\pm$~0.04 &  1.62~$\pm$~0.40 & 364/400 & 54 & 1.1~$\times$~10$^{-13}$ (1) & 57.1 & 4.61& 1.7--57\\

87GB 150844.6+571424 & 1.54~$\pm$~0.08 & $<$0.75 & 123/97 & 0 & --- &  10.9 & 0.546& 1.6--53\\ 

PKS~2000$-$330 & 1.75~$\pm$~0.07  & $<$0.55 & 146/171 & 3 & 6.3~$\times$~10$^{-2}$ (1) & 14.2 & 1.27& 1.4--48\\

PKS~2126$-$15 & 1.44~$\pm$~0.01  & 1.19~$\pm$~0.10 & 1378/1223 & 479 & 1.0~$\times$~10$^{-99}$ (1) & 133.7 & 12.05& 1.3--43\\

PKS~2149$-$306 & 1.47~$\pm$~0.01  & $<$0.03 & 1044/958 & 0 & --- & 32.3 & 3.48& 1.0--33\\

{\bf RADIO QUIET}\\

LBQS~0053$-$0134 & 2.38$^{+0.64}_{-0.45}$ & $<$1.54 & 24/33 & 1 & 0.249 (1) & 0.1 & 0.036& 0.9--31\\

PSS~J0248+1802 & 2.04$^{+0.13}_{-0.12}$ & $<$0.72 & 62/47 & 0 & --- & 4.9 & 0.2& 1.6--54\\

SDSS~1030+0524 & 2.14$^{+0.33}_{-0.18}$ & $<$6.01 & 18/14 & 0 & --- & 0.8 & 0.015& 2.2--73\\

HE 1104$-$1805 & 1.90~$\pm$~0.11 & $<$0.13 & 42/44 & 0 & --- & 1.0$^{**}$ & 0.207& 1.0--33\\

SBS~1107+487 & 1.96$^{+0.53}_{-0.33}$ & $<$1.92 & 49/59 & 1 & 0.28 (1) & 0.9 & 0.111 & 1.2--40\\

Q~1442+2931   &  1.90~$\pm$~0.07 & $<$0.16 & 120/124 & 0 & --- & 1.9 & 0.300 & 1.1--36\\

BR~2237$-$0607 & 2.15$^{+0.34}_{-0.25}$ & $<$3.77 & 18/20 & 1 & 0.30 (1) & 0.4 & 0.101& 1.7--56\\

{\bf BAL}\\

Q~0000$-$263 & 2.07~$\pm$~0.13 & $<$0.82 & 75/85 & 0 & --- & 1.7 & 0.091& 1.5--51\\

APM 08279+5255$^{a}$  & 1.91$^{+0.07}_{-0.06}$ & 5.50$^{+0.83}_{-0.76}$ & 221/199 & 198 & 3.6~$\times$~10$^{-32}$ (1) & 14.2$^{**}$ & 0.7 & 1.5--49\\

RX~J0911.4+0551 & 1.63$^{+0.20}_{-0.18}$ & 2.15$^{+1.56}_{-1.19}$ & 63/64 & 11 & 1.4~$\times$~10$^{-3}$ (1) & 1.5$^{**}$ & 0.2& 1.1--38\\

Q 1246-057 & 1.86$^{+0.32}_{-0.26}$ & 1.36$^{+1.33}_{-0.92}$ & 61/68 & 7 &  6.8~$\times$~10$^{-3}$ (1) & 0.2& 0.04& 1.0--32\\

CSO 0755 &  1.90~$\pm$~0.08 & 1.52~$\pm$~0.38 & 112/143 & 62 & 2.3~$\times$~10$^{-15}$ (1) & 2.5 & 0.191&1.2--39\\

SBS 1542+541 & 1.80$^{+0.15}_{-0.14}$ & $<$1.16 & 72/57 & 8 & 1.5~$\times$~10$^{-2}$ (1) & 0.7 & 0.2& 1.0--37\\

\hline
\end{tabular}
\end{center}
\end{table*}

\begin{figure*}
\begin{center}
\includegraphics[clip,width=2.5cm,angle=-90]{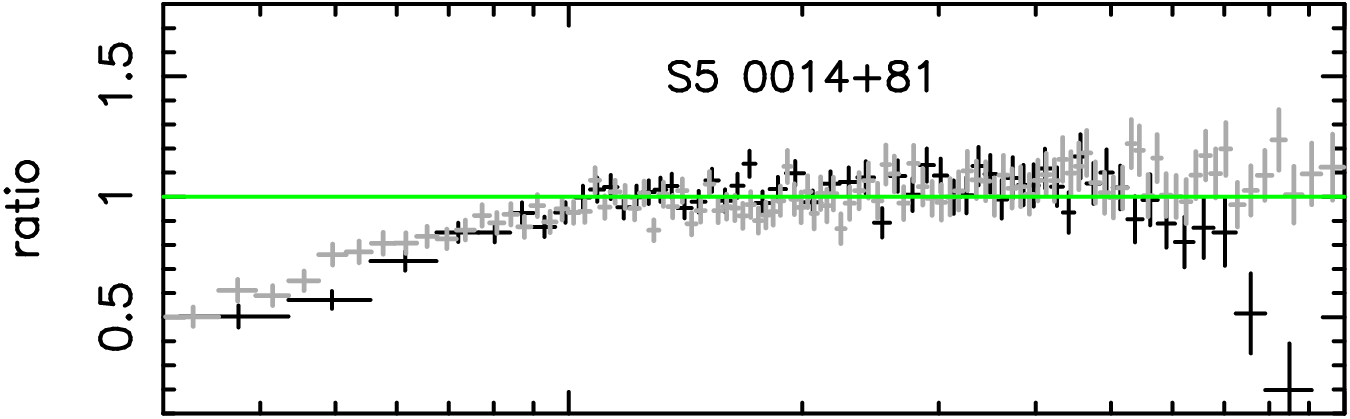}\hspace*{0.45cm}
\includegraphics[clip,width=2.5cm,angle=-90]{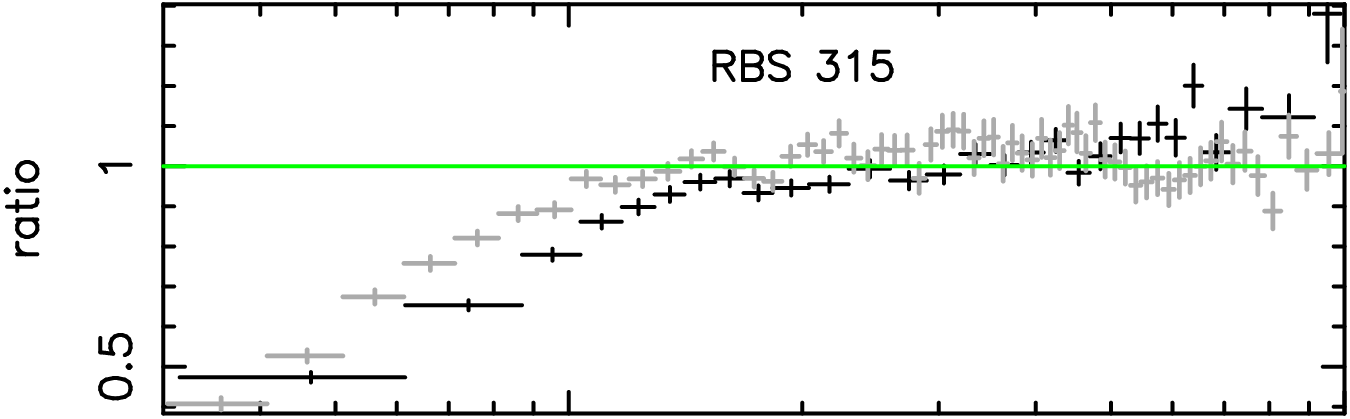}
\includegraphics[clip,width=2.5cm,angle=-90]{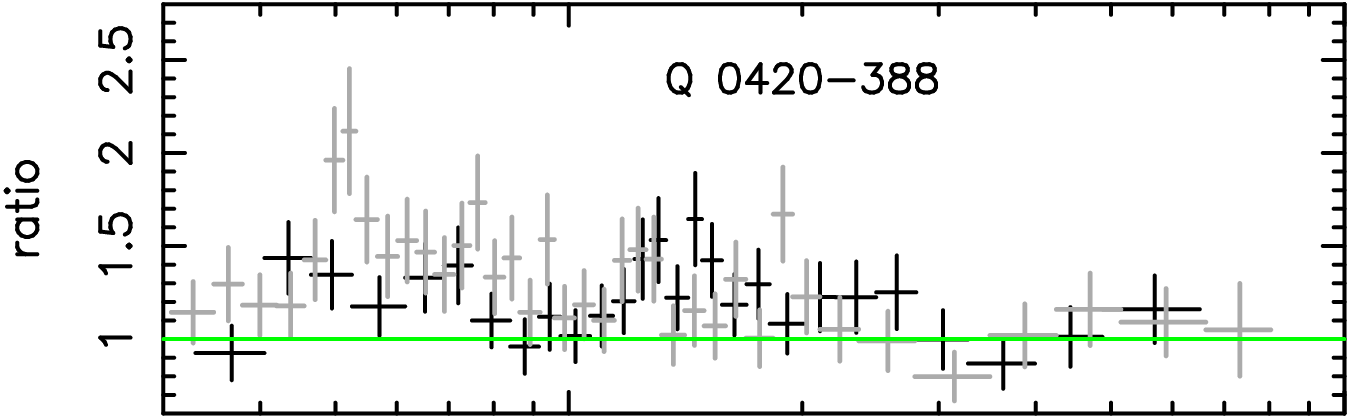}\hspace*{0.45cm}
\includegraphics[clip,width=2.5cm,angle=-90]{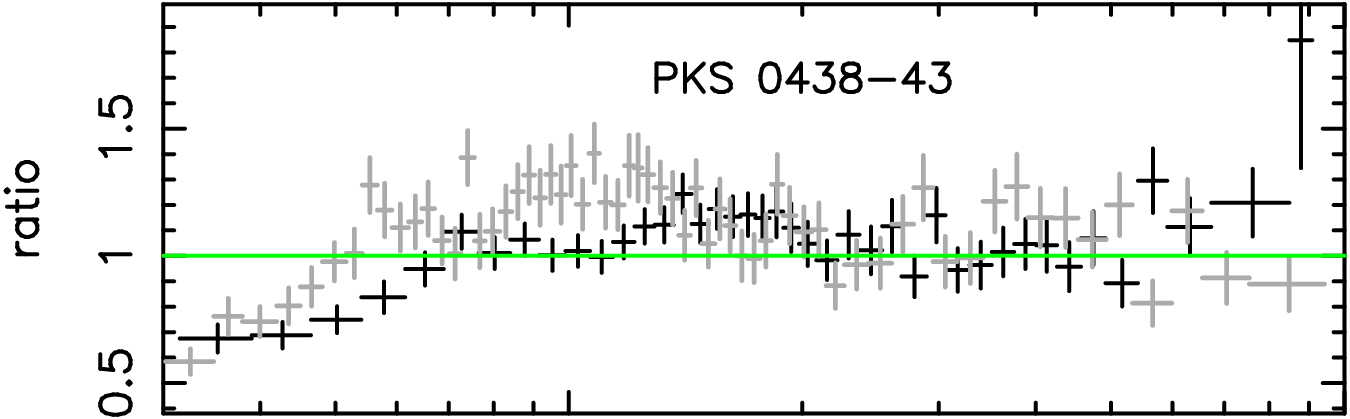}
\includegraphics[clip,width=2.5cm,angle=-90]{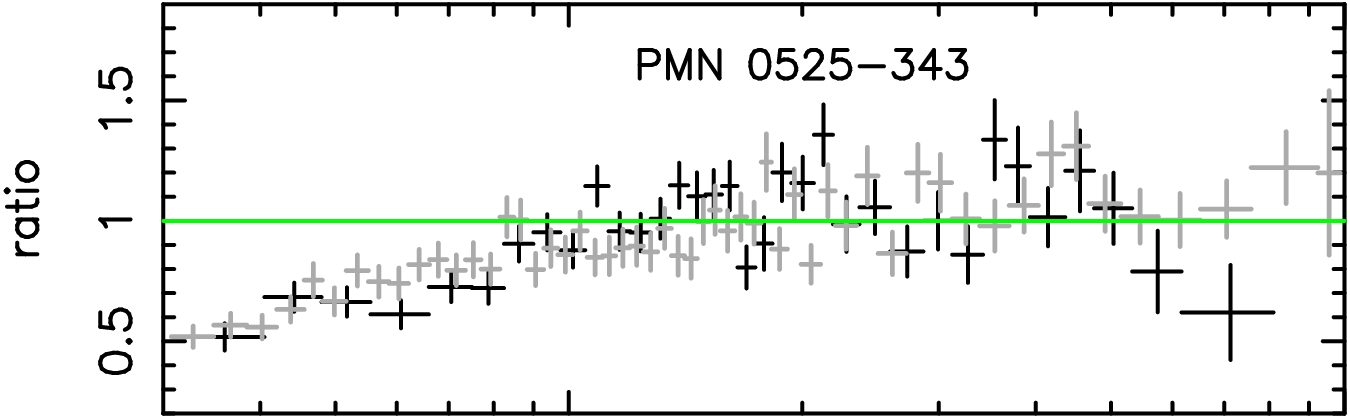}\hspace*{0.45cm}
\includegraphics[clip,width=2.5cm,angle=-90]{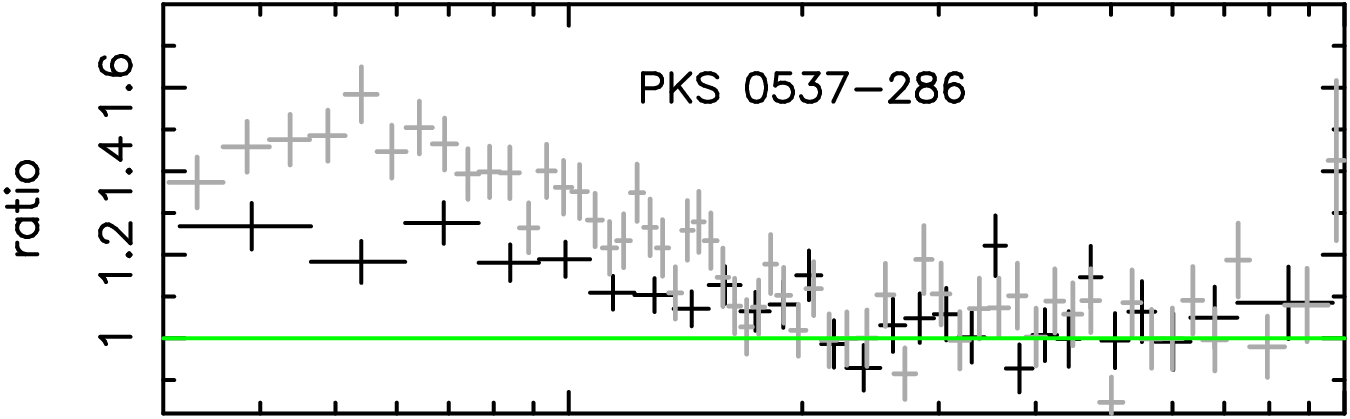}
\includegraphics[clip,width=2.5cm,angle=-90]{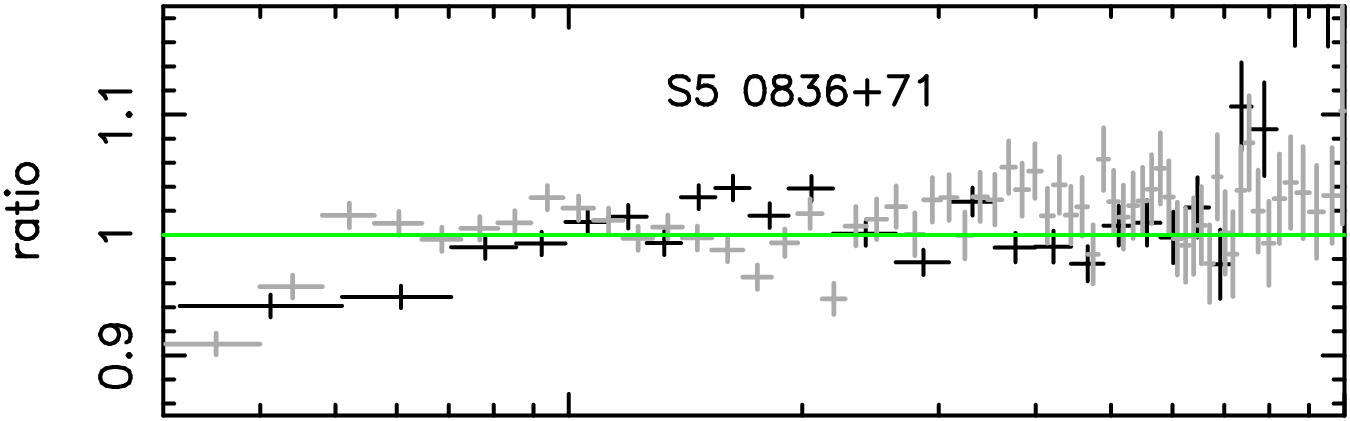}\hspace*{0.45cm}
\includegraphics[clip,width=2.5cm,angle=-90]{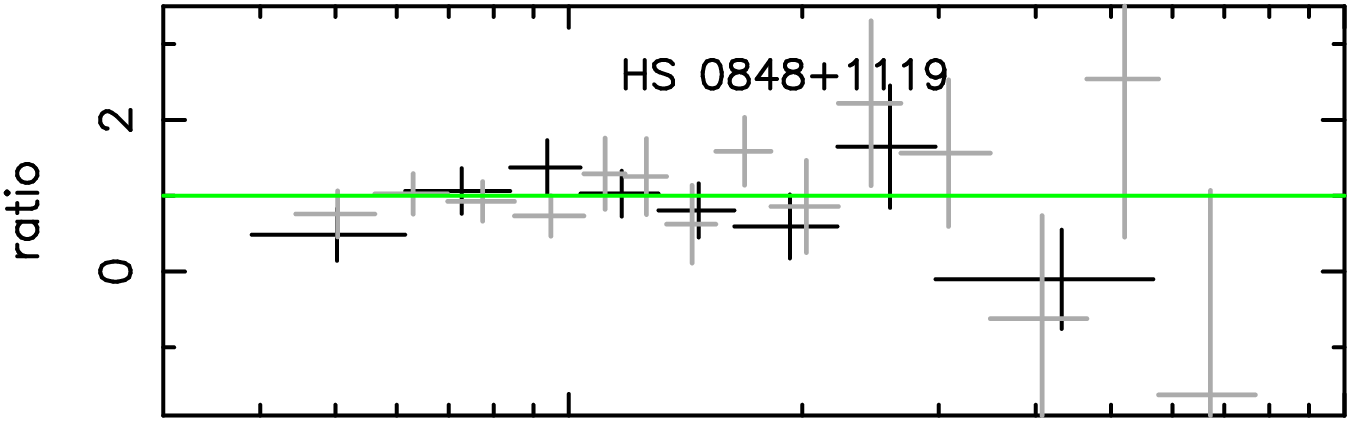}
\includegraphics[clip,width=2.5cm,angle=-90]{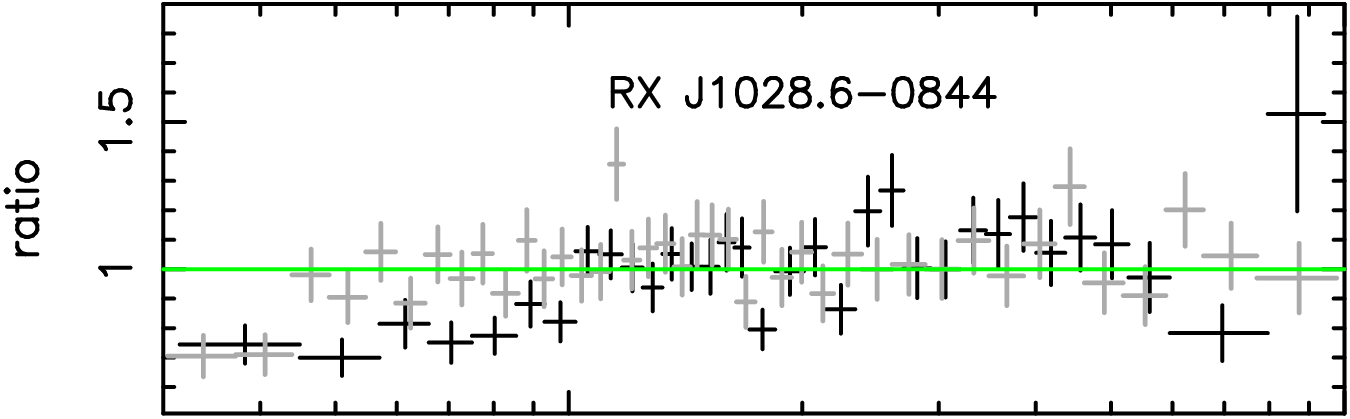}\hspace*{0.45cm}
\includegraphics[clip,width=2.5cm,angle=-90]{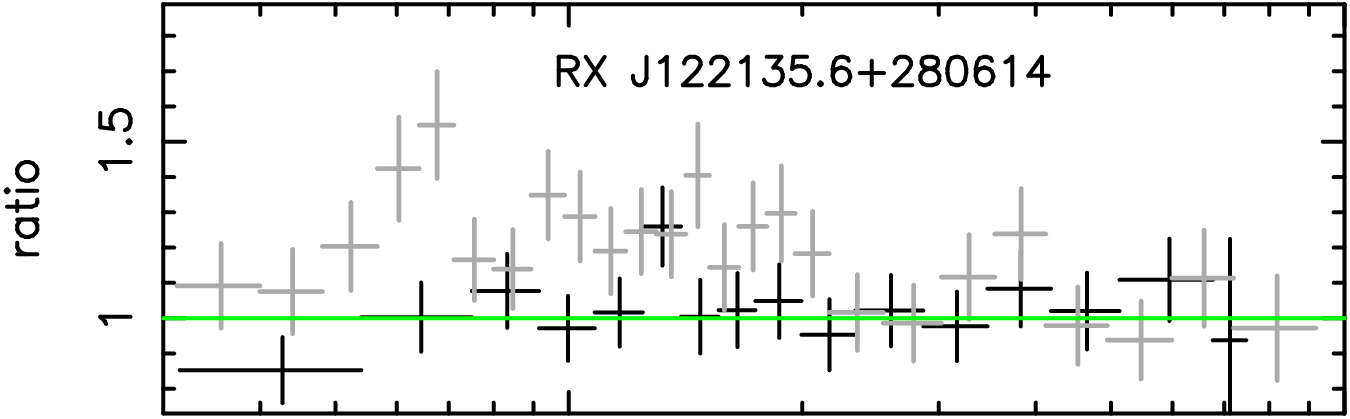}
\includegraphics[clip,width=2.5cm,angle=-90]{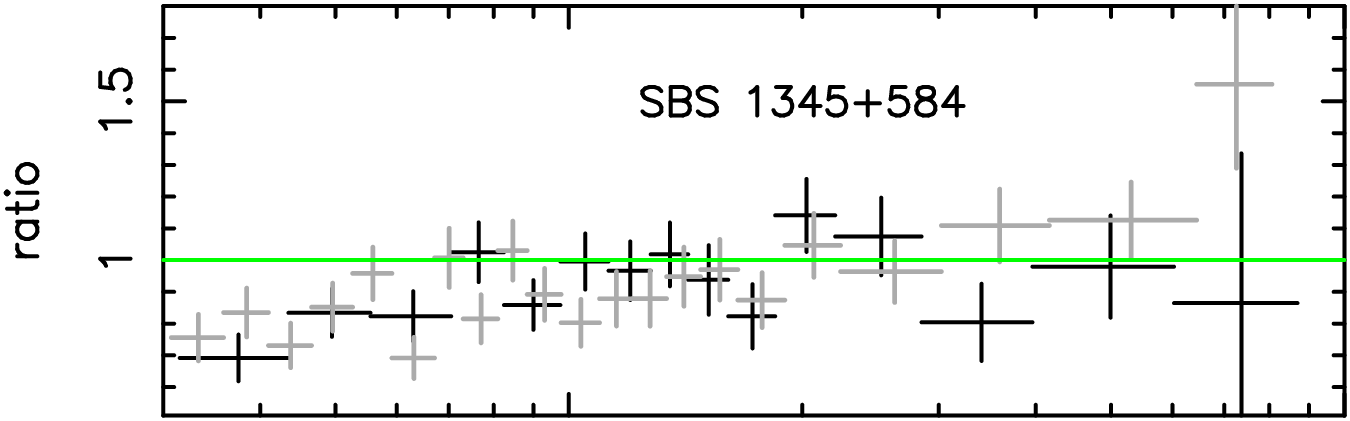}\hspace*{0.45cm}
\includegraphics[clip,width=2.5cm,angle=-90]{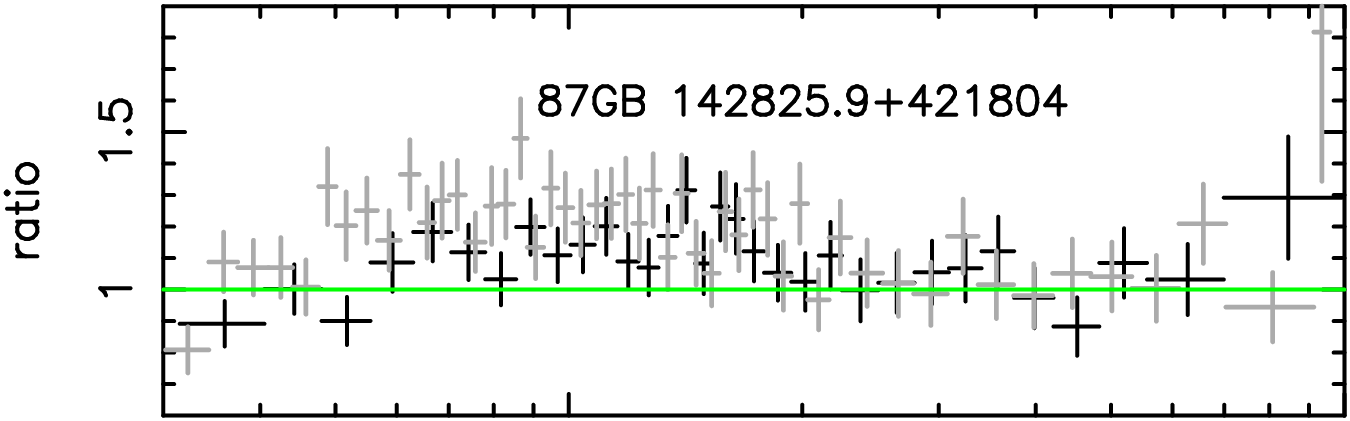}
\includegraphics[clip,width=2.5cm,angle=-90]{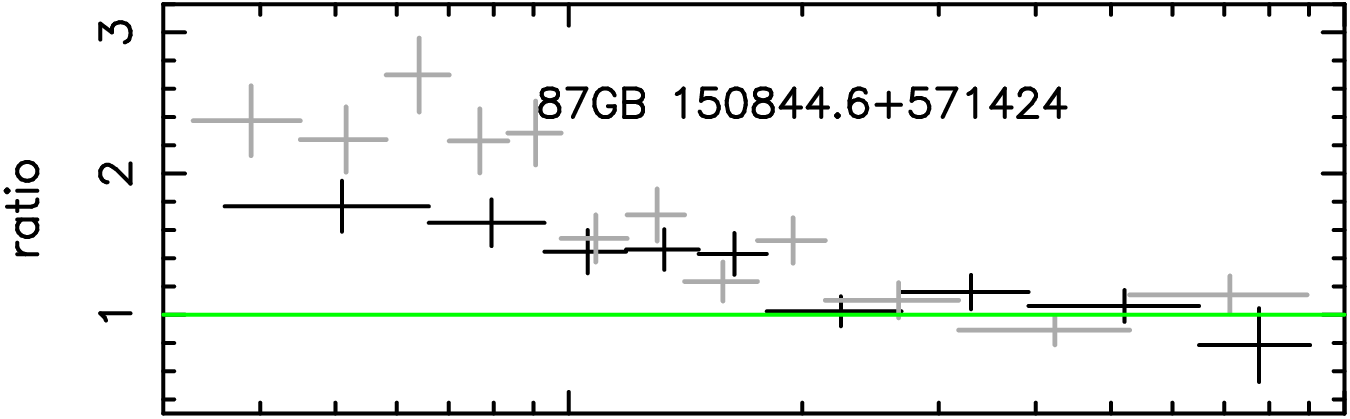}\hspace*{0.45cm}
\includegraphics[clip,width=2.5cm,angle=-90]{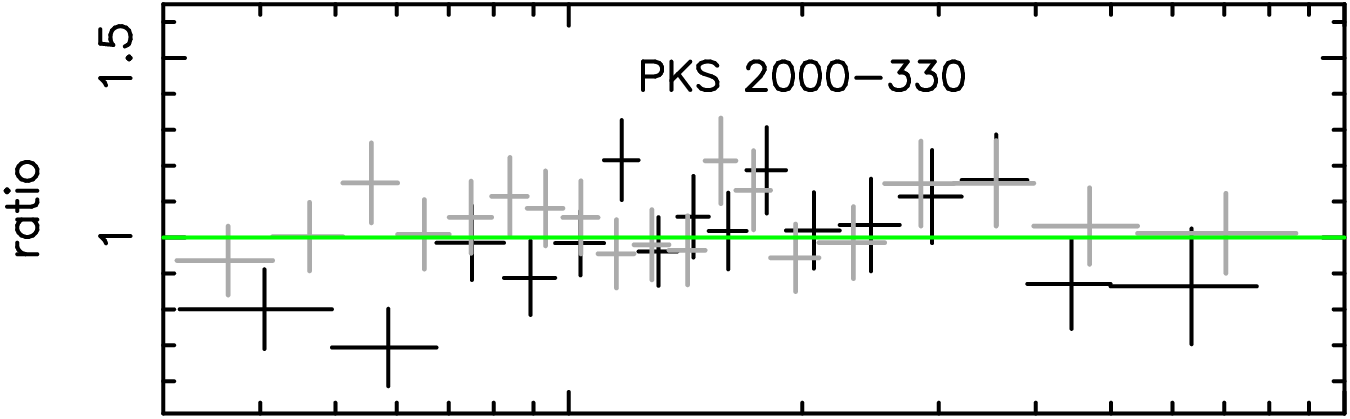}
\includegraphics[clip,width=3.3cm,angle=-90]{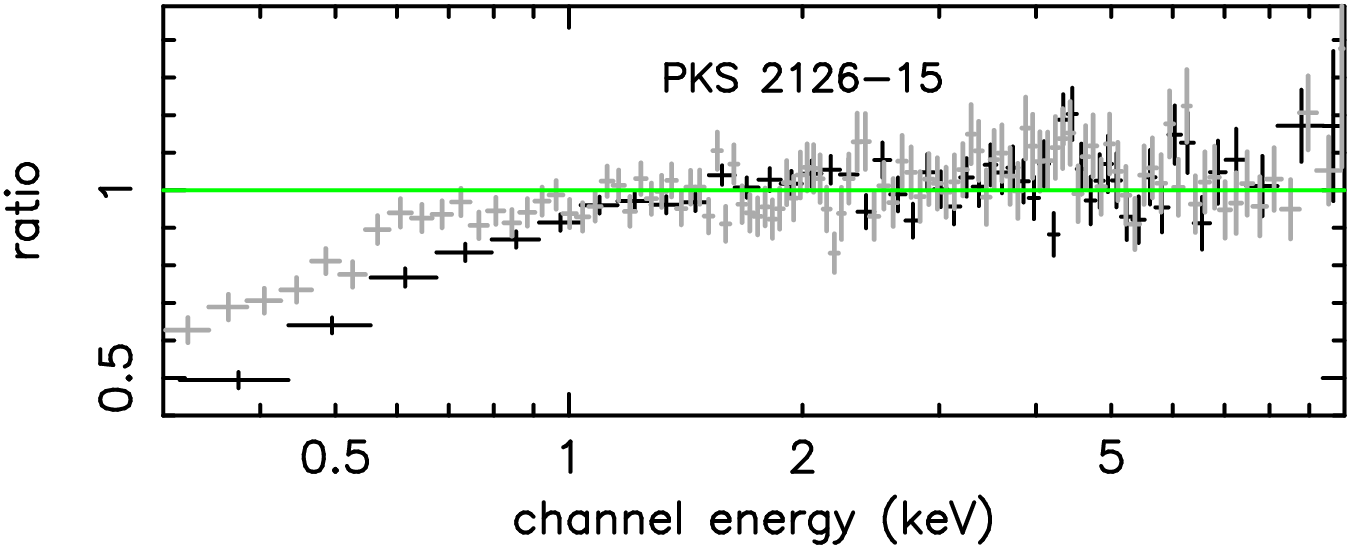}\hspace*{0.45cm}
\includegraphics[clip,width=3.25cm,angle=-90]{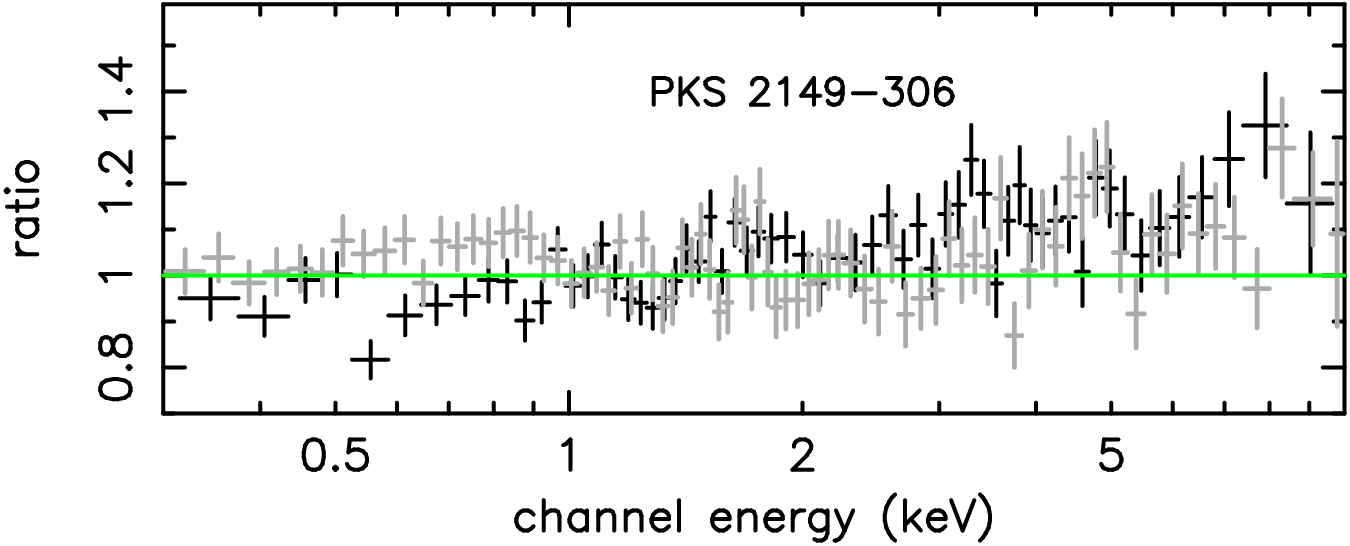}
\caption{Power-law fits over the observed-frame 2--10~keV band are extrapolated over 0.3--10~keV, for the RLQs. In most cases this reveals evidence for curvature, generally in the sense of absorption at lower energies. MOS data are shown in black, PN, in grey.}
\label{broad_rlq}
\end{center}
\end{figure*}

\begin{figure*}
\begin{center}
\includegraphics[clip,width=2.5cm,angle=-90]{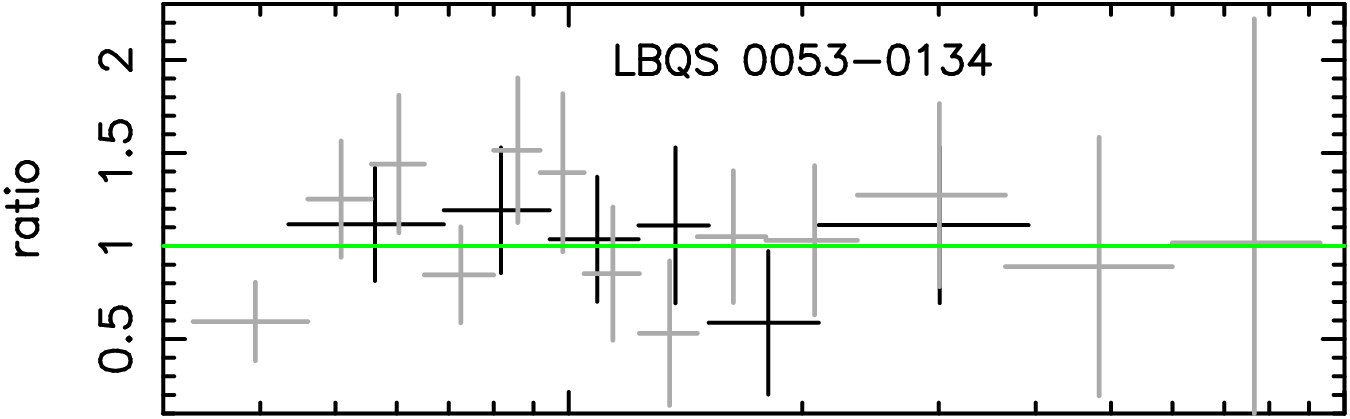}
\includegraphics[clip,width=2.5cm,angle=-90]{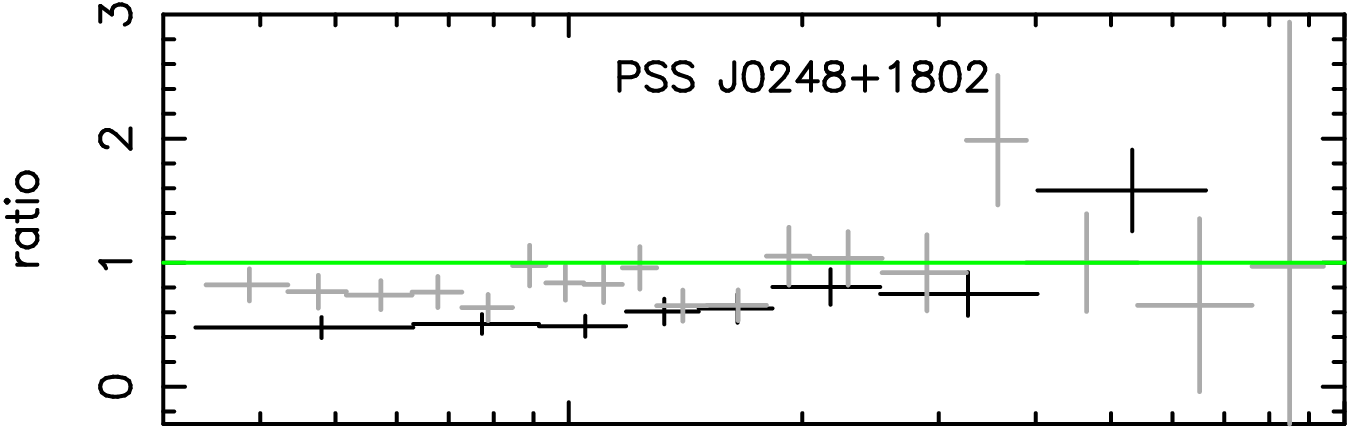}\hspace*{0.45cm}
\includegraphics[clip,width=2.5cm,angle=-90]{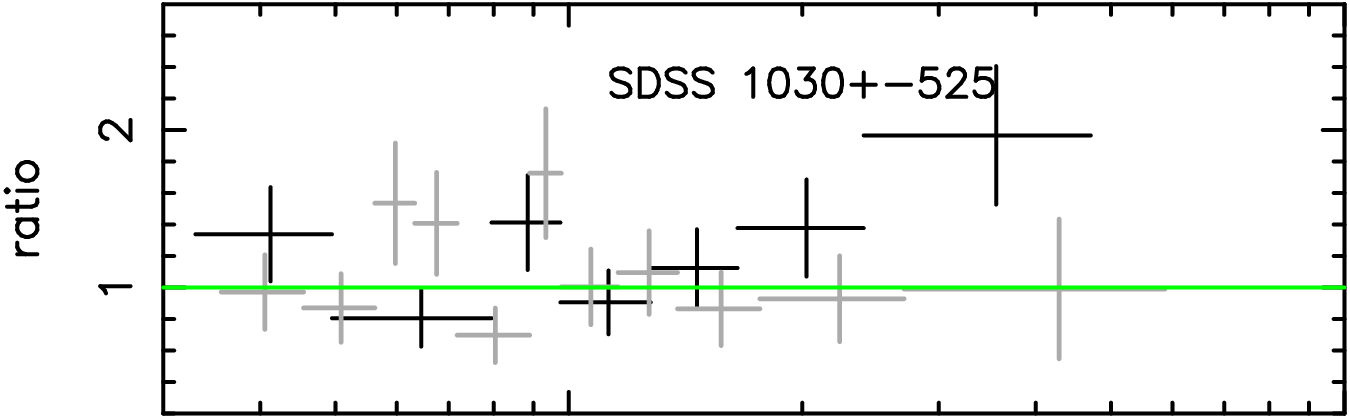}
\includegraphics[clip,width=2.5cm,angle=-90]{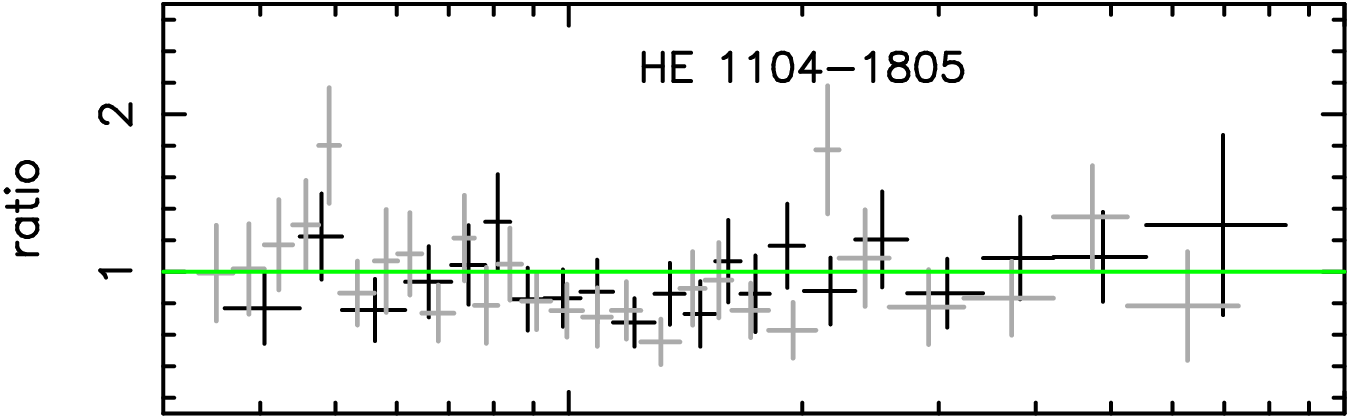}\hspace*{0.45cm}
\includegraphics[clip,width=2.5cm,angle=-90]{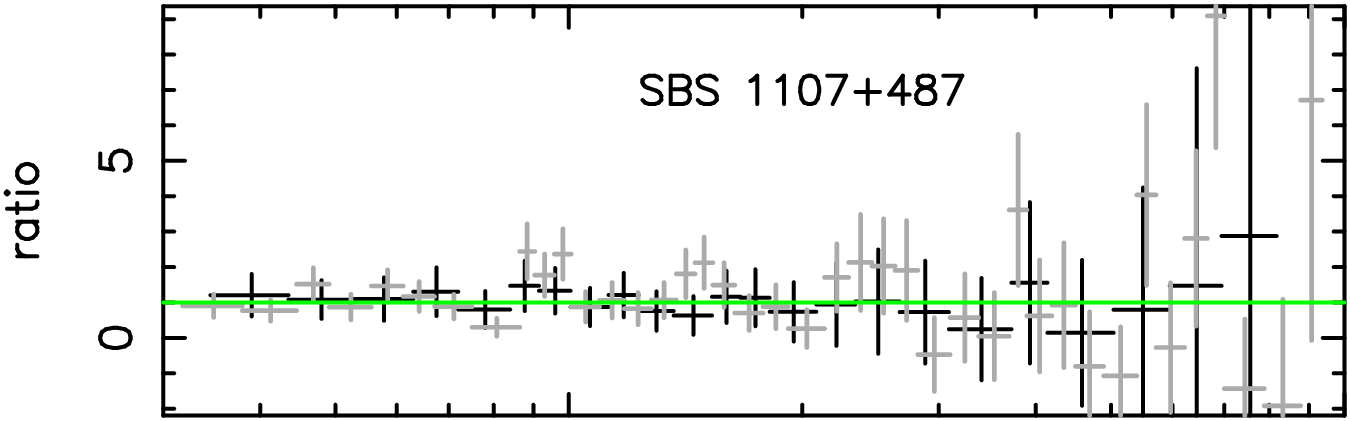}
\includegraphics[clip,width=3.25cm,angle=-90]{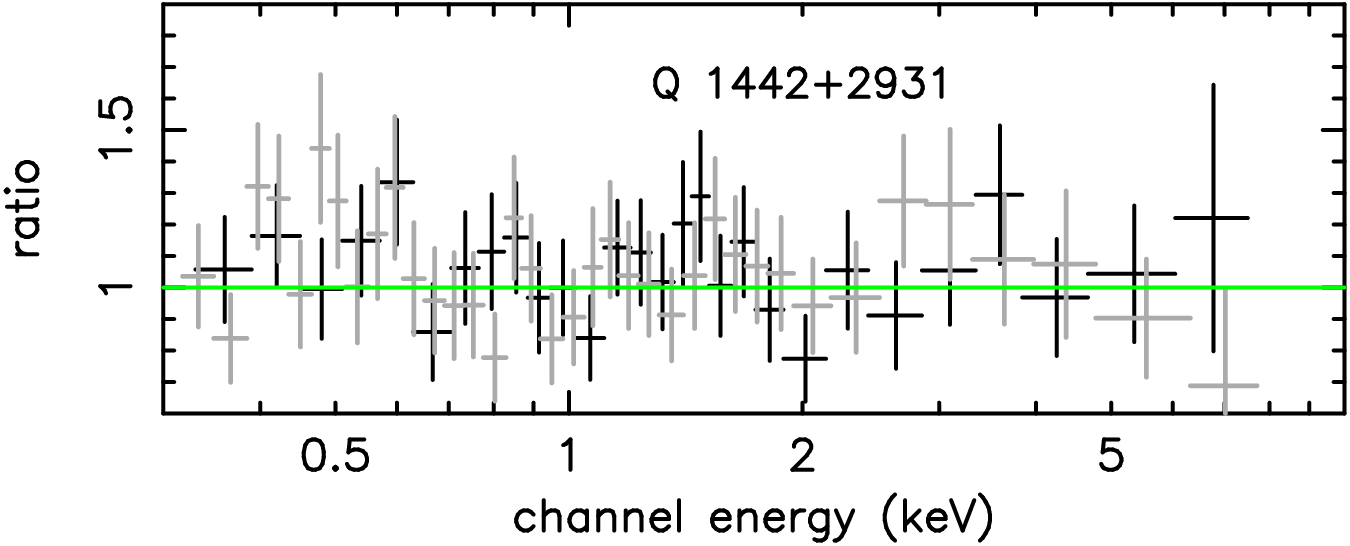}\hspace*{0.45cm}
\includegraphics[clip,width=3.25cm,angle=-90]{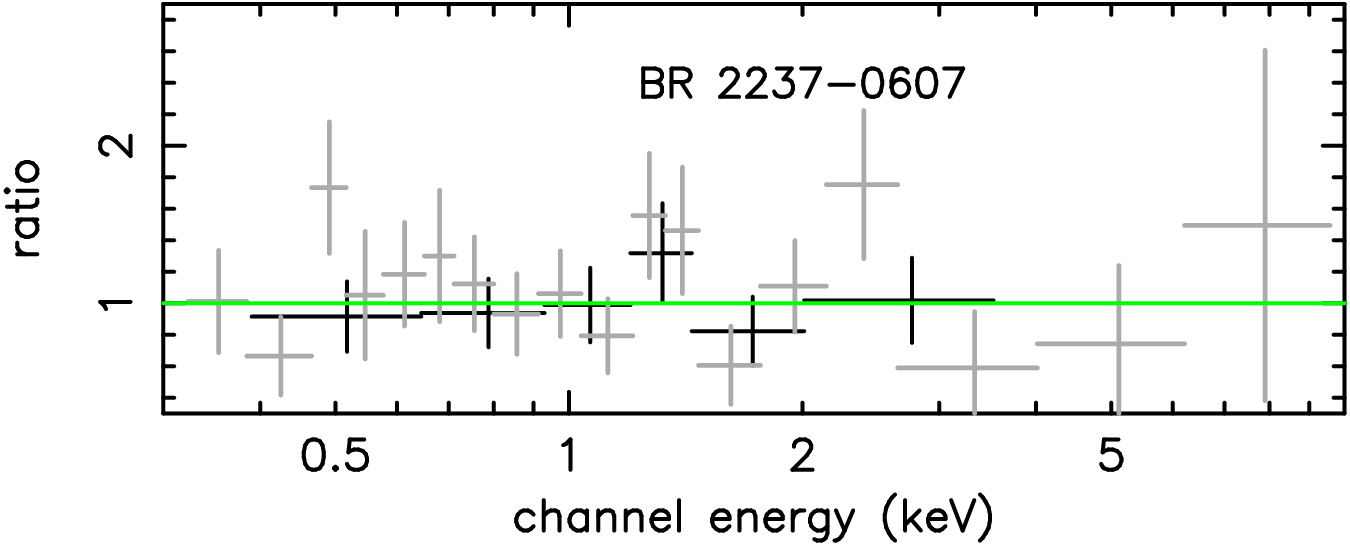}
\caption{As for Figure~\ref{broad_rlq}, but for the RQQs.}
\label{broad_rqq}
\end{center}
\end{figure*}

\begin{figure*}
\begin{center}
\includegraphics[clip,width=2.5cm,angle=-90]{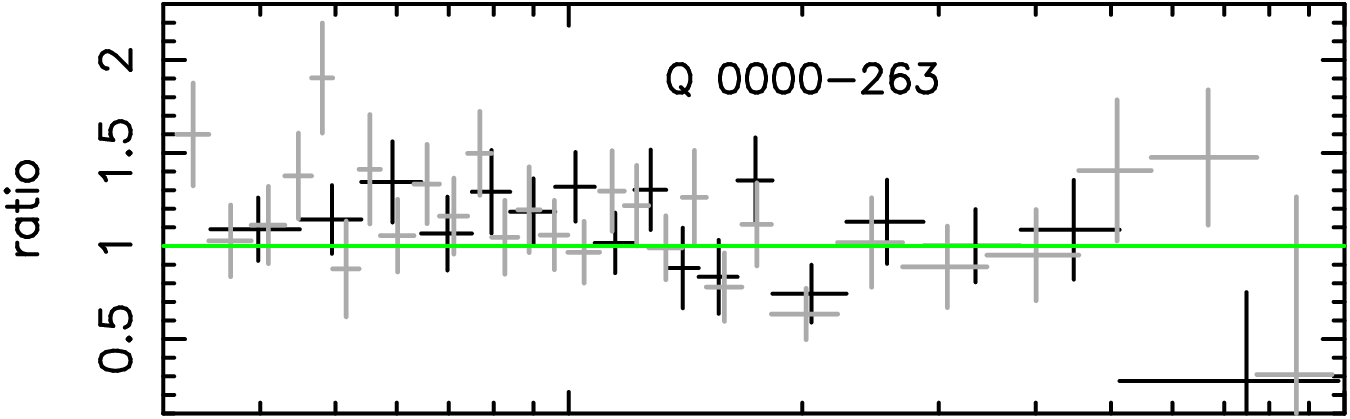}\hspace*{0.45cm}
\includegraphics[clip,width=2.5cm,angle=-90]{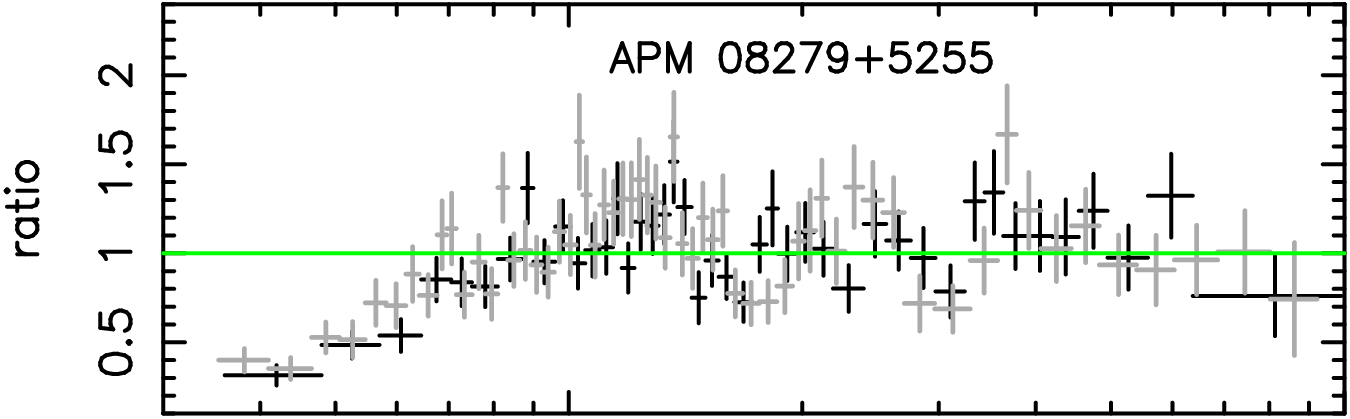}
\includegraphics[clip,width=2.5cm,angle=-90]{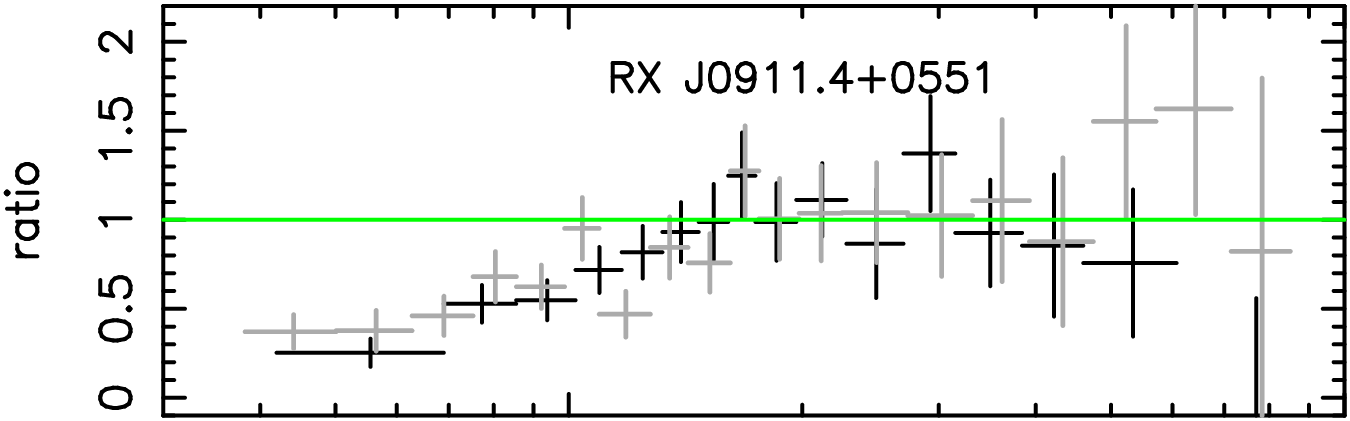}\hspace*{0.45cm}
\includegraphics[clip,width=2.5cm,angle=-90]{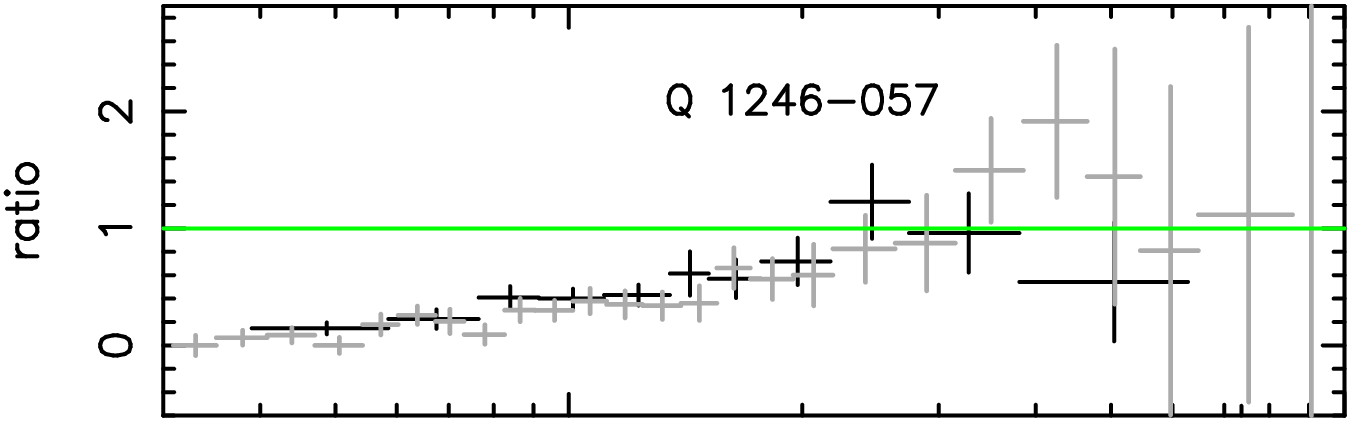}
\includegraphics[clip,width=3.25cm,angle=-90]{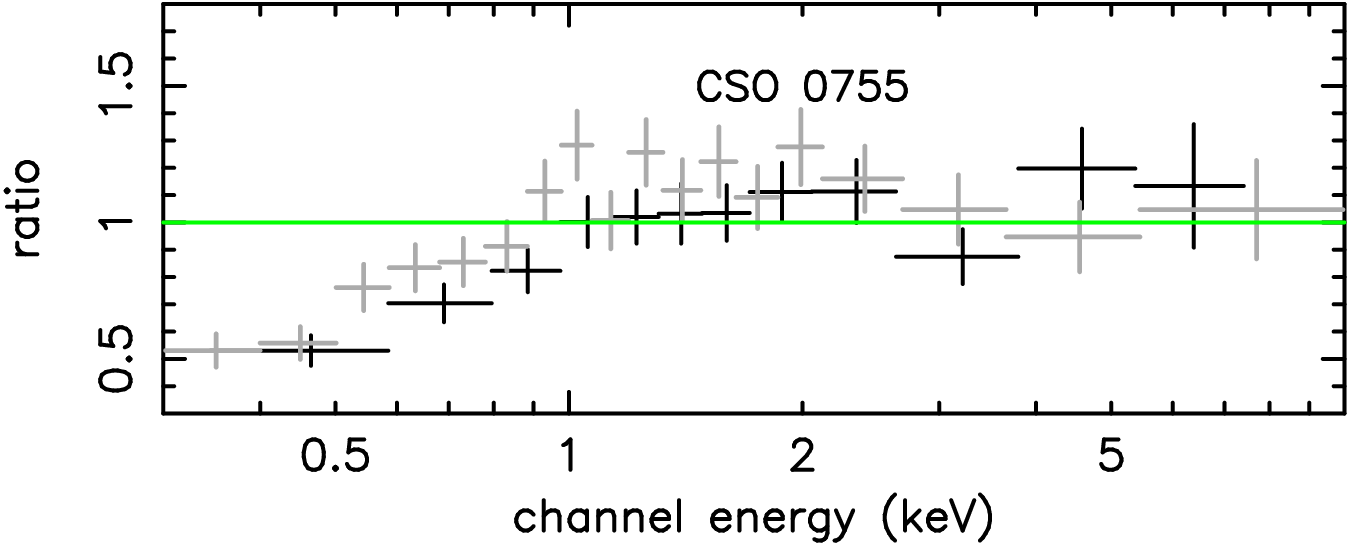}\hspace*{0.45cm}
\includegraphics[clip,width=3.25cm,angle=-90]{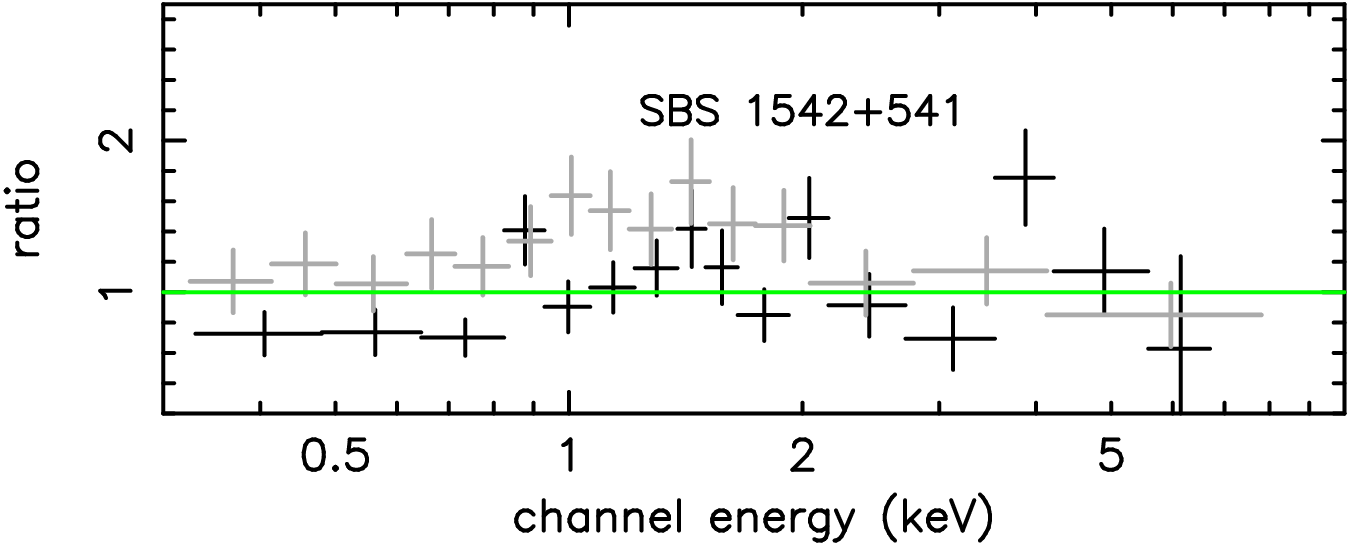}

\caption{As for Figure~\ref{broad_rlq}, but for the BAL QSOs. Note the absorption edge just below 2~keV (observed) in the spectrum of APM~08279+5255.}
\label{broad_bal}
\end{center}
\end{figure*}

For each QSO, the rest-frame 2--10~keV power law fit was extrapolated over the full bandpass of {\it XMM-Newton}. Figure~\ref{broad_rlq} shows the ratio plots formed by fitting a simple power-law over the {\it observed} frame 2--10~keV, and extrapolating over 0.3--10~keV for the RLQs, while Figures~\ref{broad_rqq} and~\ref{broad_bal} shows the results for the RQQs and BAL QSOs. From these three figures, it can be seen that the majority of the spectra showing curvature appear convex in shape, indicating the possible presence of excess absorption. Some, however, are more concave, implying they may be better fitted by a broken power-law model.

\subsection{Complex continuum fits}

\begin{figure*}
\begin{center}

\includegraphics[clip,width=4.0cm,angle=-90]{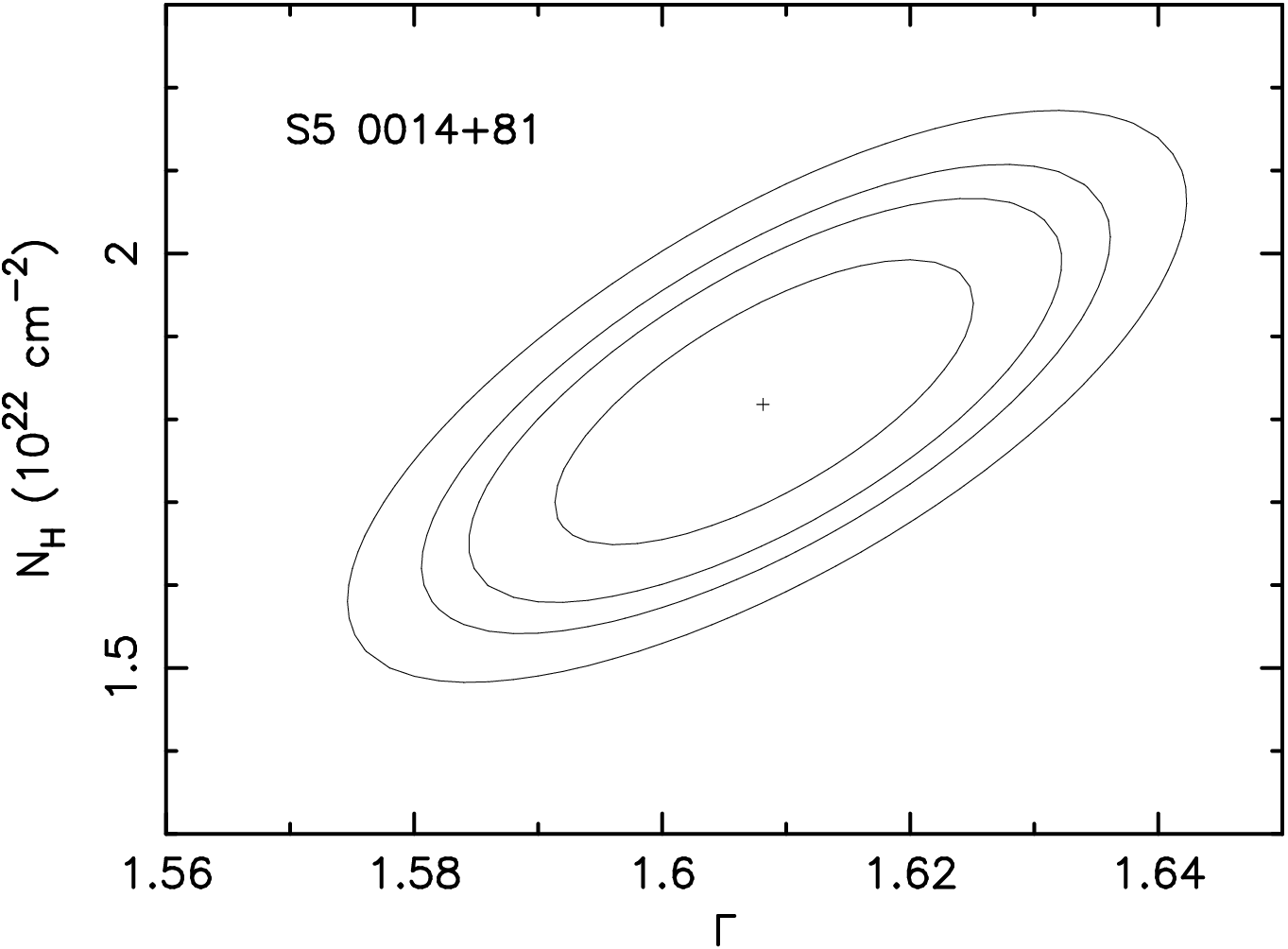}\hspace*{0.5cm}
\includegraphics[clip,width=4.0cm,angle=-90]{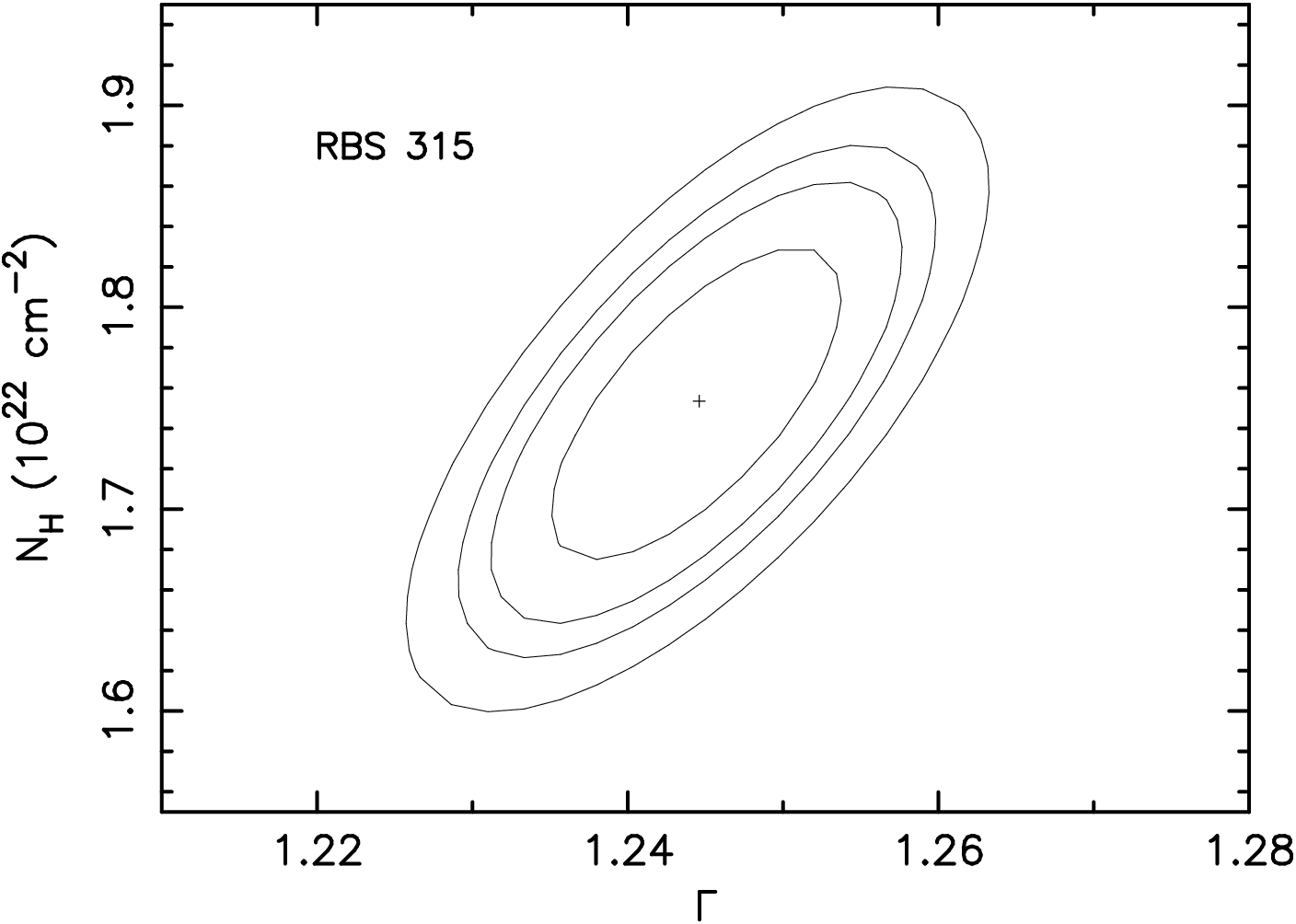}\hspace*{0.5cm}
\includegraphics[clip,width=4.0cm,angle=-90]{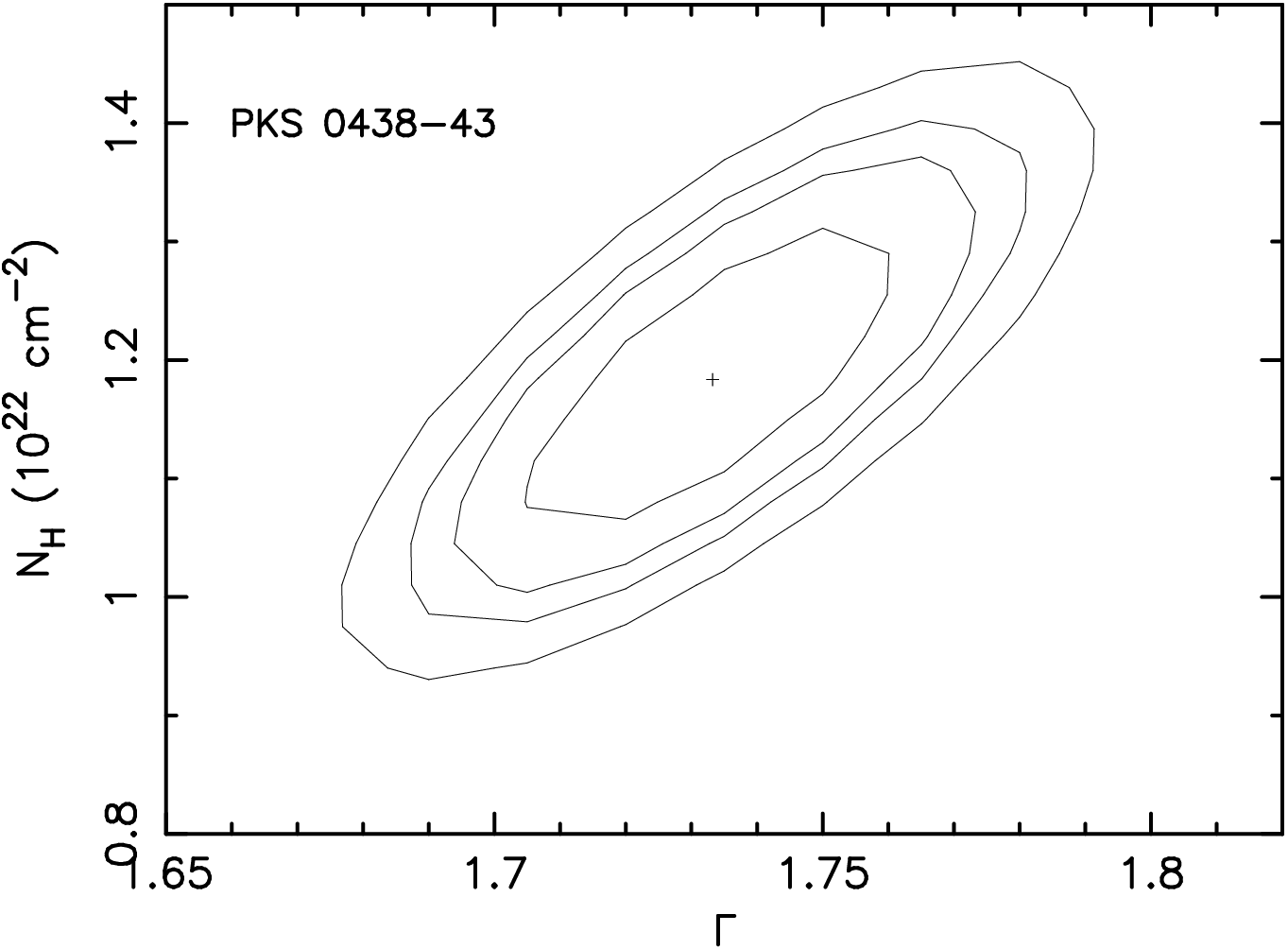}\vspace*{0.4cm}
\includegraphics[clip,width=4.0cm,angle=-90]{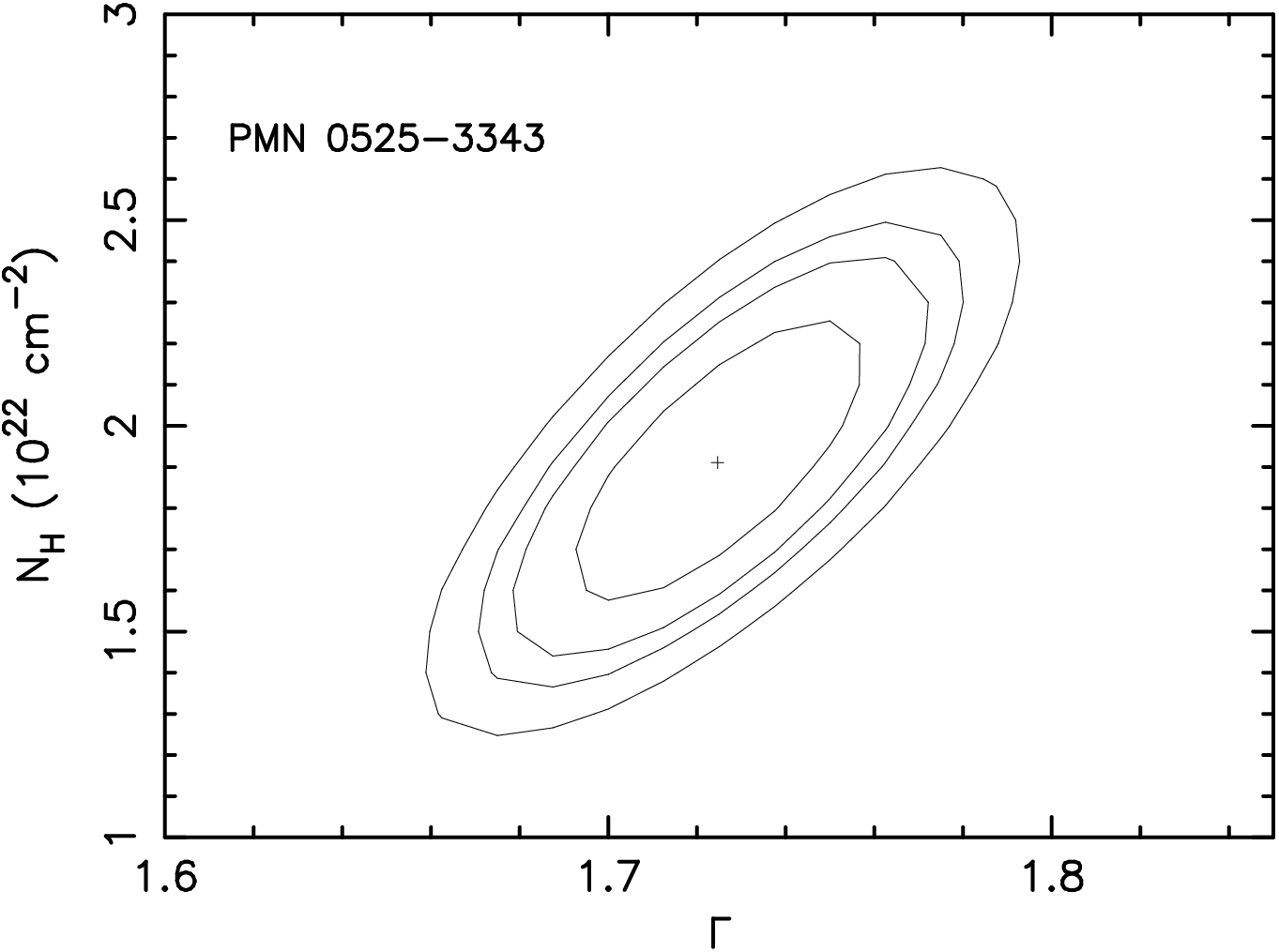}\hspace*{0.5cm}
\includegraphics[clip,width=4.0cm,angle=-90]{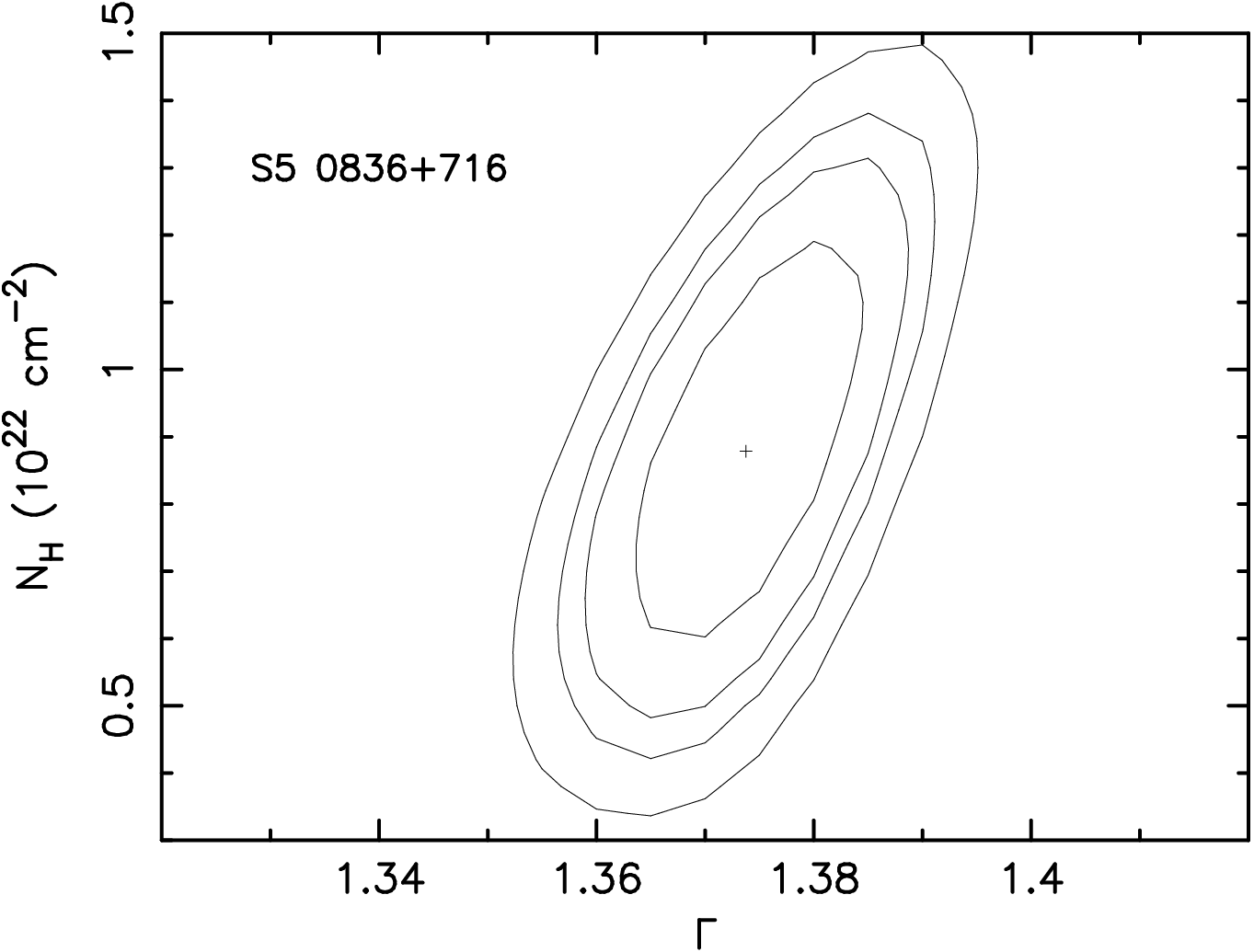}\hspace*{0.5cm}
\includegraphics[clip,width=4.0cm,angle=-90]{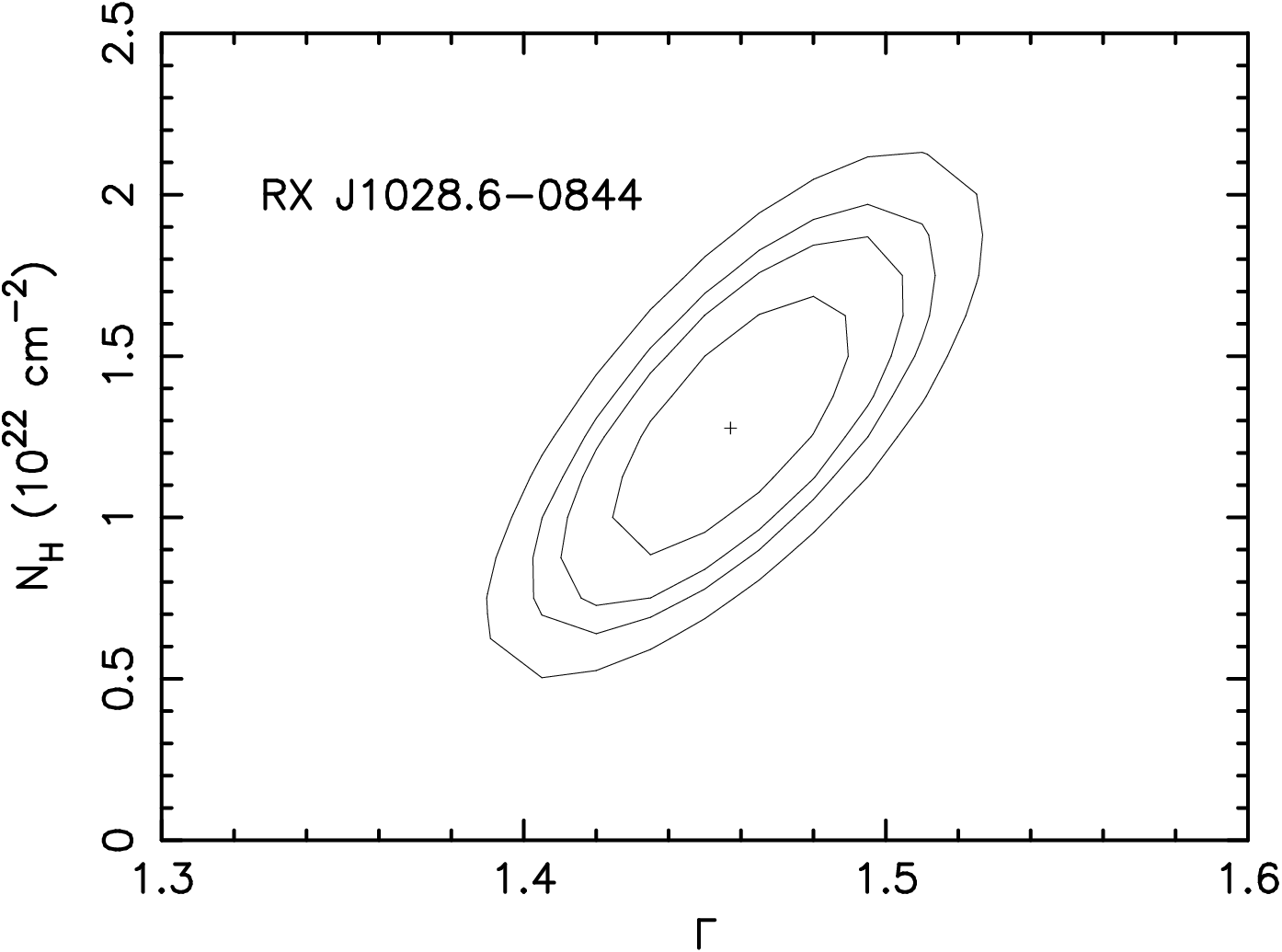}\vspace*{0.4cm}
\includegraphics[clip,width=4.0cm,angle=-90]{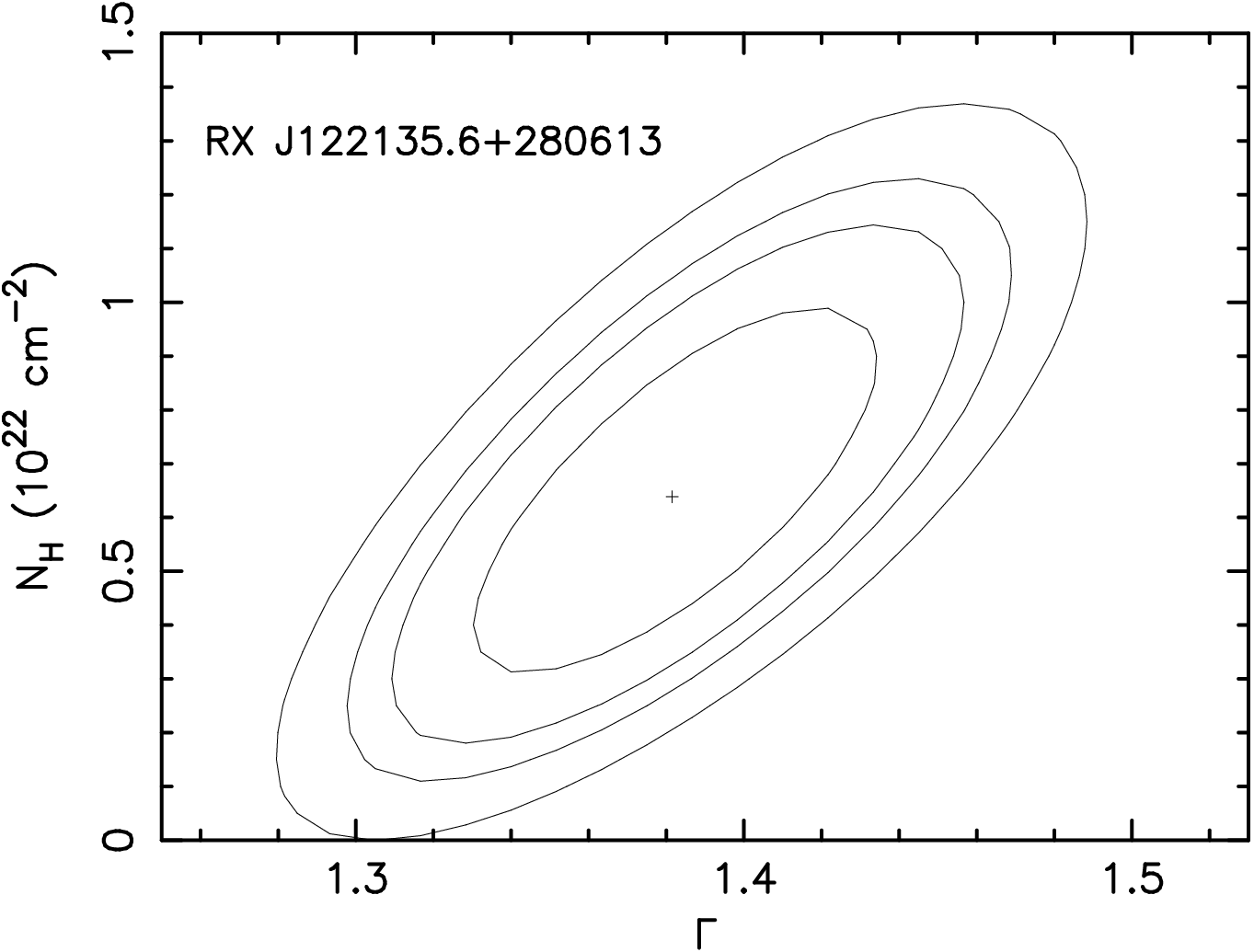}\hspace*{0.5cm}
\includegraphics[clip,width=4.0cm,angle=-90]{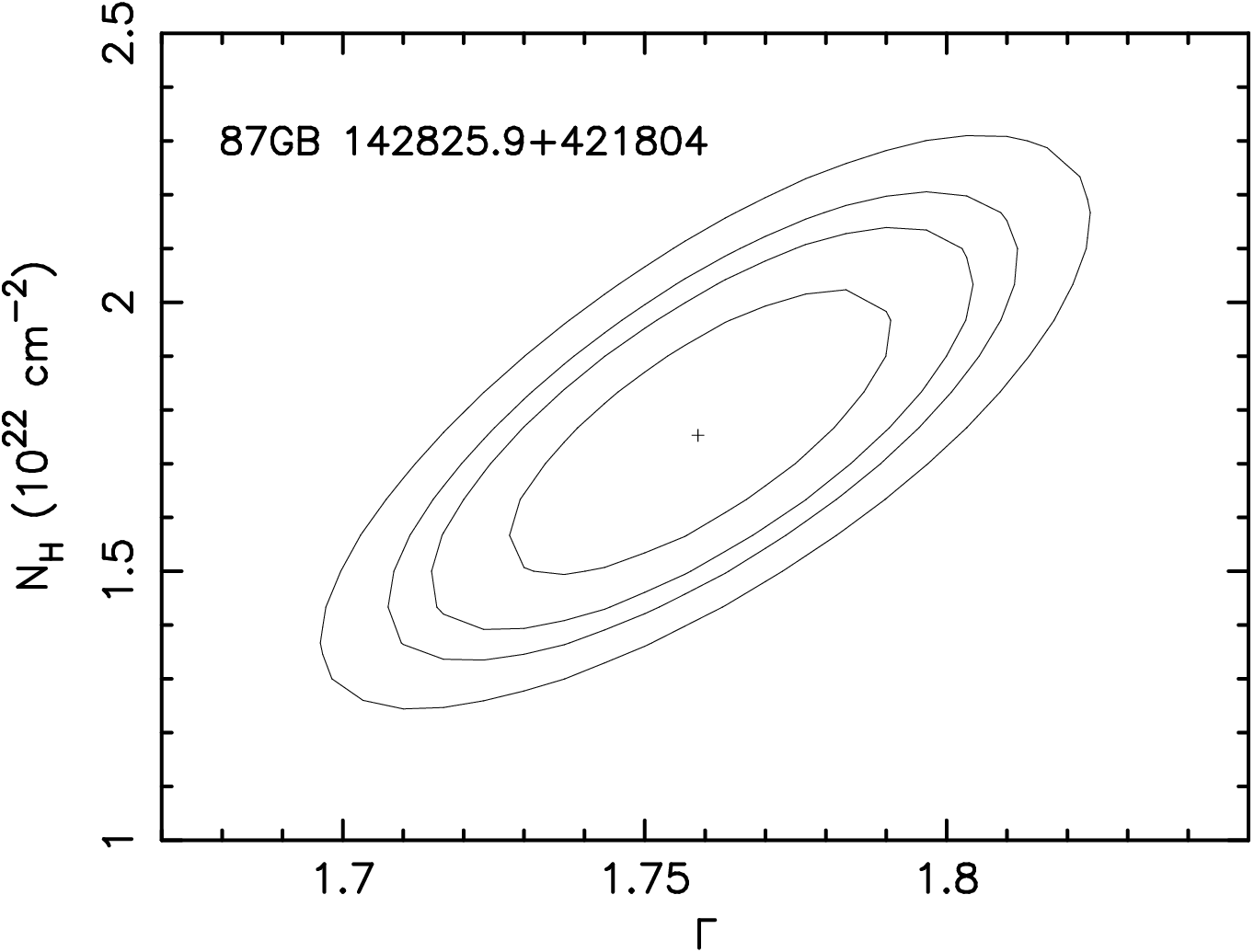}\hspace*{0.5cm}
\includegraphics[clip,width=4.0cm,angle=-90]{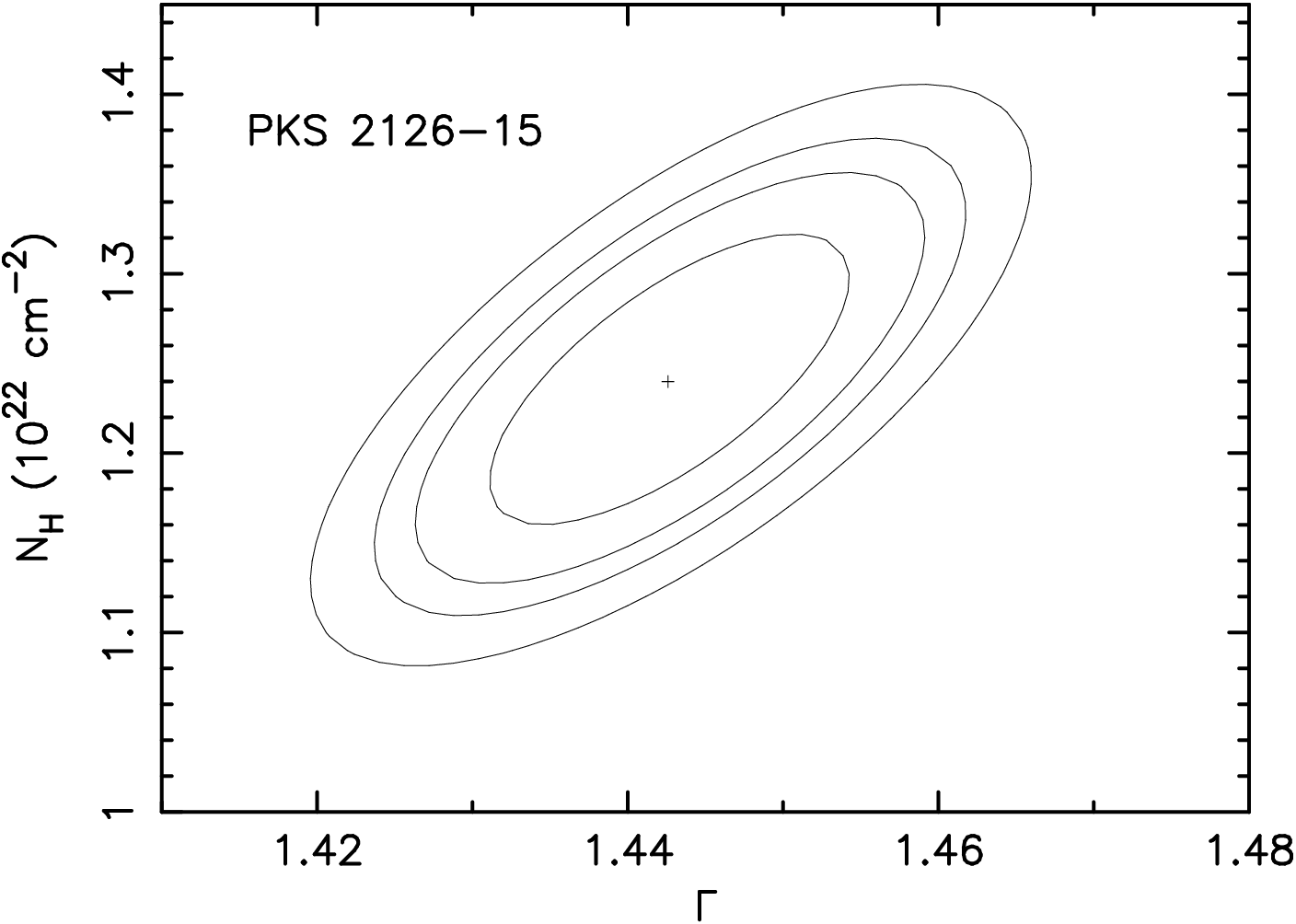}\vspace*{0.4cm}
\includegraphics[clip,width=4.0cm,angle=-90]{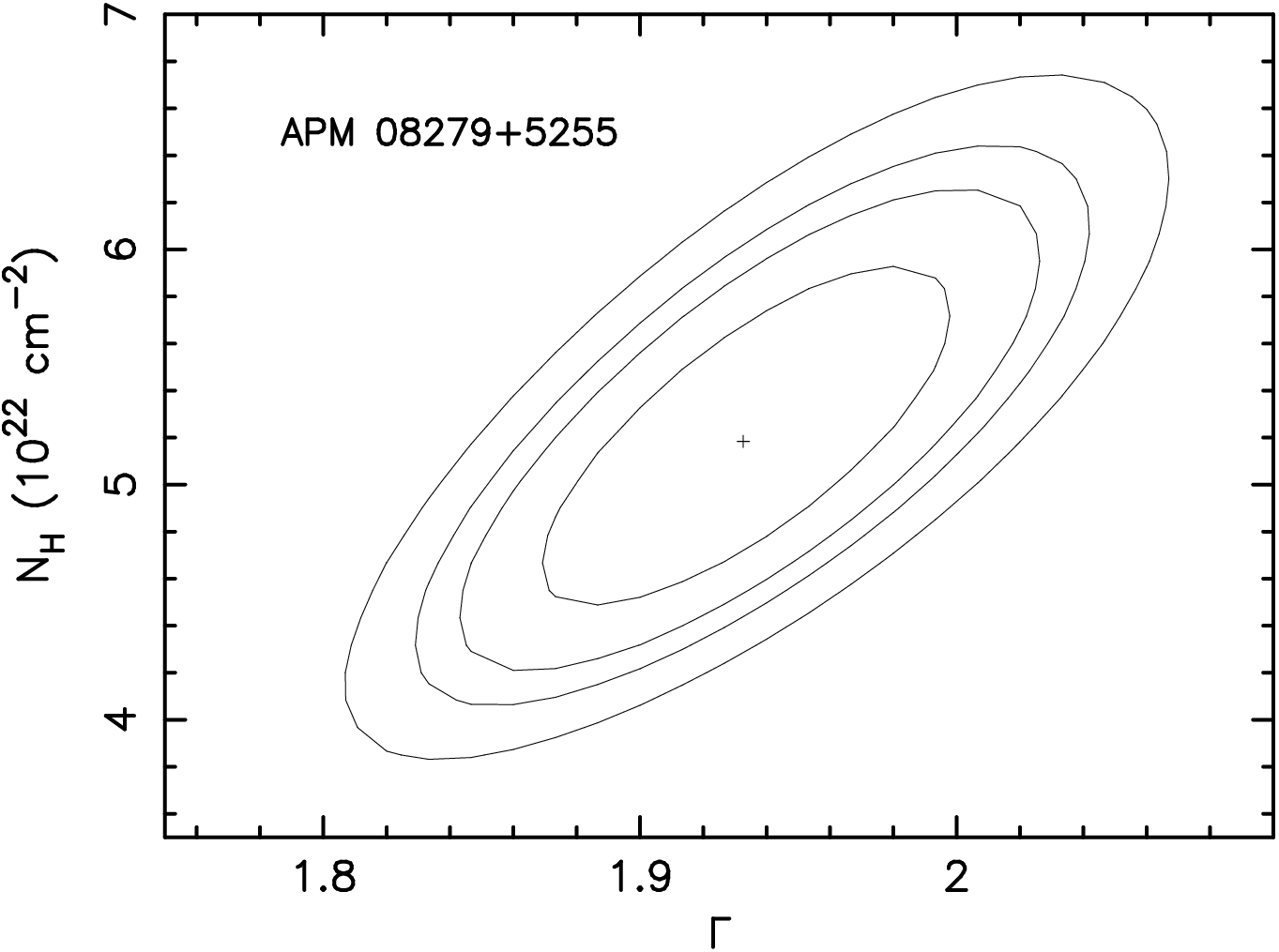}\hspace*{0.5cm}
\includegraphics[clip,width=4.0cm,angle=-90]{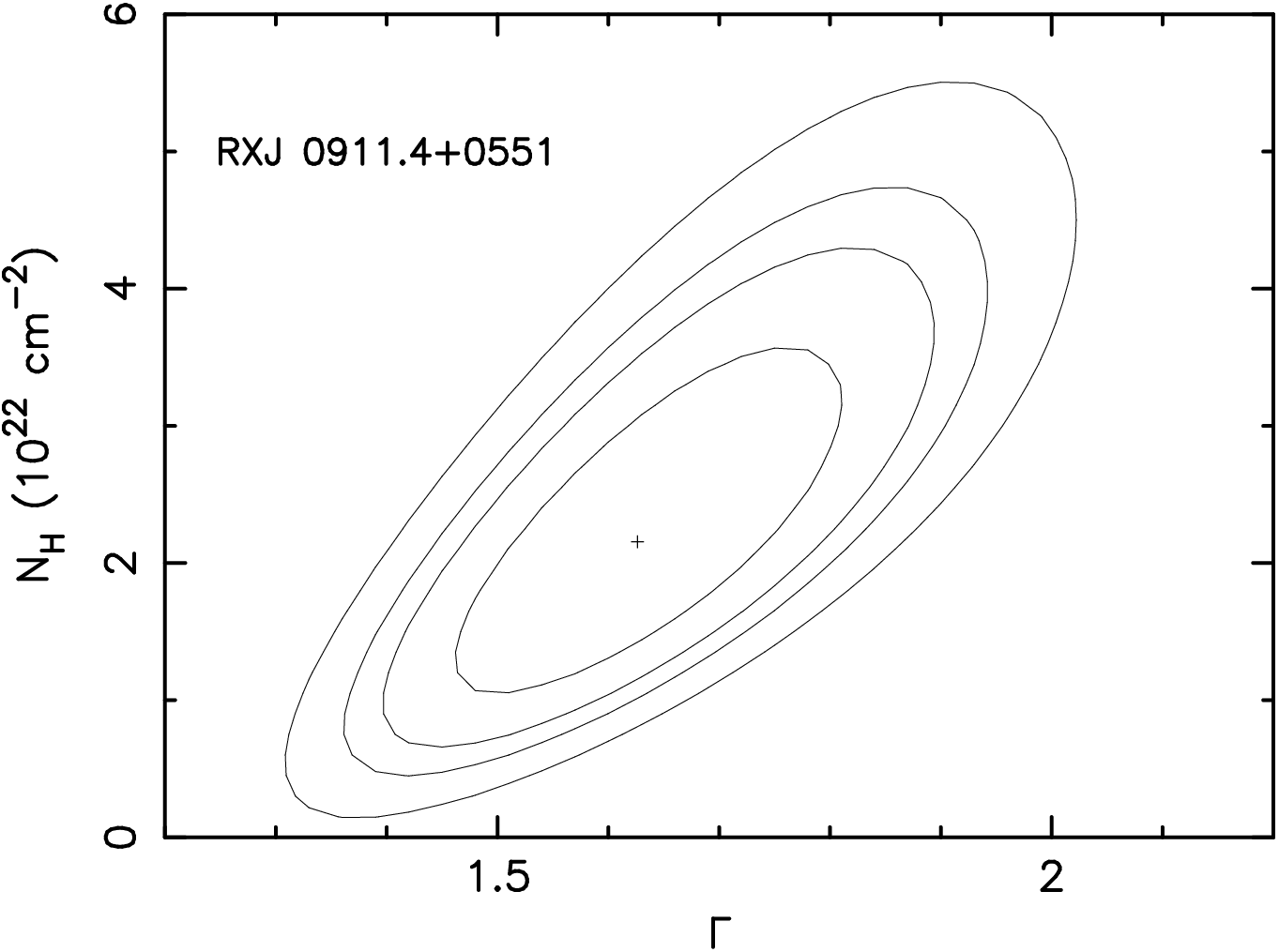}\hspace*{0.5cm}
\includegraphics[clip,width=4.0cm,angle=-90]{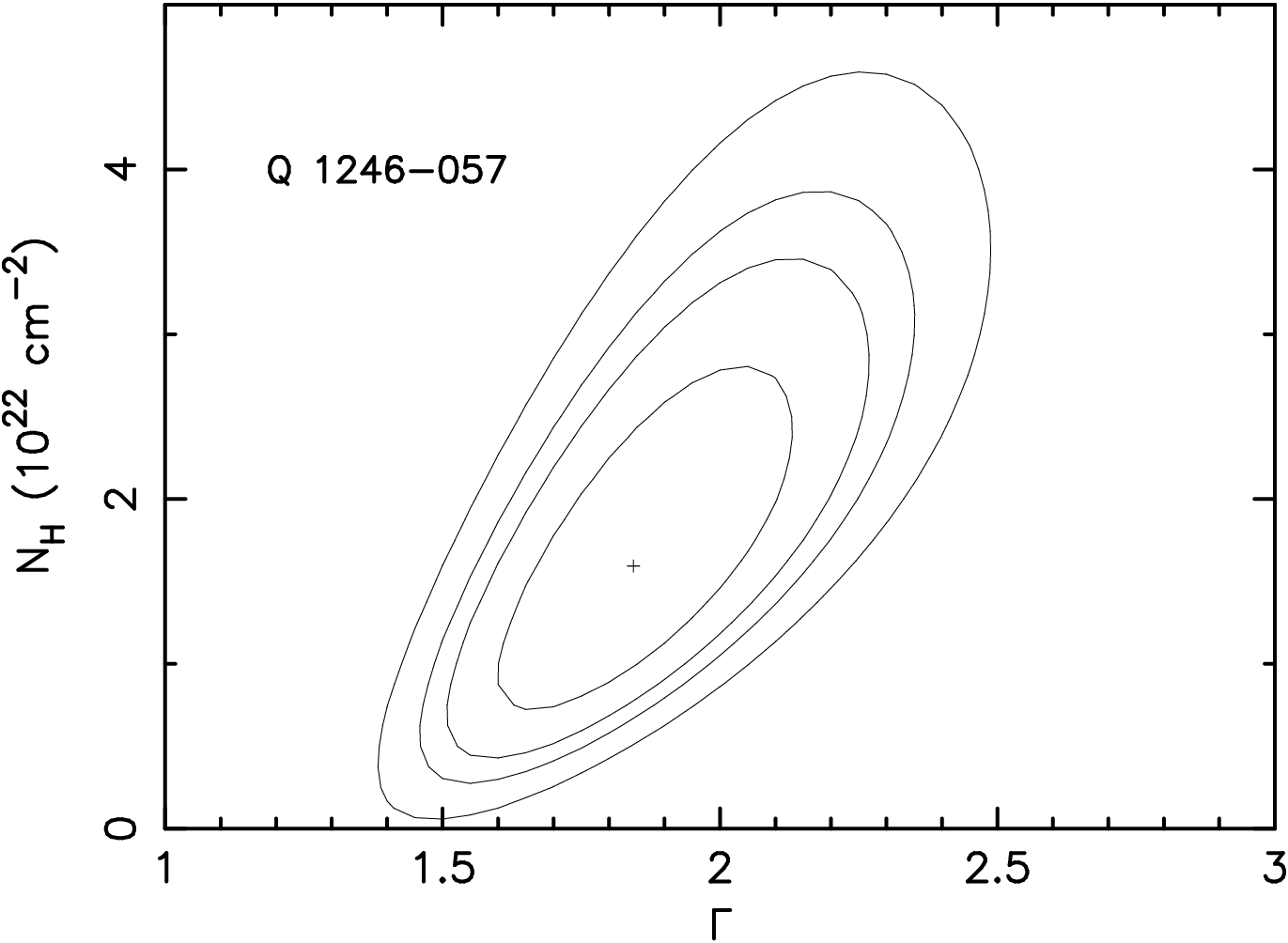}\vspace*{0.4cm}
\includegraphics[clip,width=4.0cm,angle=-90]{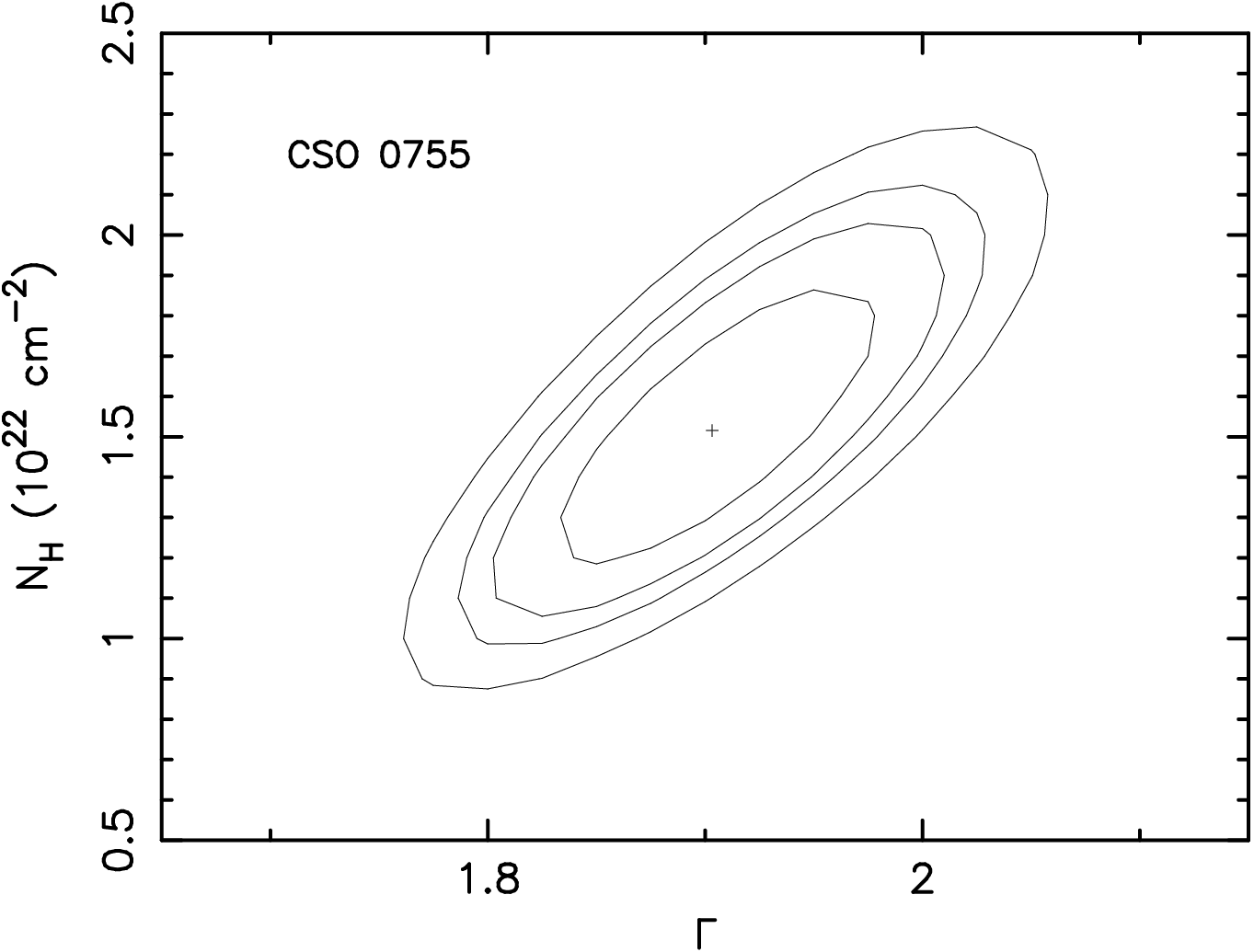}
\caption{68, 90, 95 and 99~per~cent confidence contours for the intrinsic absorbing column density and the photon index, $\Gamma$, for the QSOs where excess absorption is significant.}
\label{cont}
\end{center}
\end{figure*}

Many of the RLQs showed evidence for excess absorption in their rest-frames, between $\sim$1--2~$\times$~10$^{22}$~cm$^{-2}$ (Table~\ref{fits}). Upper limits are given for the remaining objects. Using an ionised absorber model did not improve the fits further; however, the upper limits on the ionisation were not well constrained for most cases, so the presence of ionised absorption cannot be excluded. This was even the case for APM~08279+5255 where an ionised absorption edge improved the fit. Figure~\ref{cont} plots the confidence contours for the absorbing column and photon index for the objects where excess N$_{\rm H}$ is detected.

Broken power-law models were then fitted to the spectra of those QSOs with significant rest-frame absorption, fixing the N$_{\rm H}$ at the Galactic value. Also included were PKS~0537$-$286 and 87GB~150844.6+571424, since the ratio plots in Figure~\ref{broad_rlq} indicated the presence of curvature in these spectra. Using a broken power-law investigates whether what is being modelled by excess N$_{\rm H}$ is actually just the flattening of the continua at low energies, rather than real absorption. Table~\ref{curv} gives the results of these fits and compares the reduced $\chi^{2}$ values with those found from the excess  N$_{\rm H}$ fits. The final column shows which of the fits has the lower $\chi^{2}$ value. 

It was found that the spectra of S5~0014+81, RBS 315, PKS~0438$-$43, PMN~0525$-$3343 and PKS~2126$-$15 were all better fitted with excess absorption, while S5~0836+71, RX~J1028.6$-$0844, RX~J122135.6+280613, RX~J0911.4+0551 and CSO~0755 were better modelled by a broken power-law. The two models gave the same result for 87GB~142825.9+421804 and very similar values for RX~J122135.6+280613 and APM~08279+5255. Despite the statistical fit being slightly improved by using a broken power-law for these few objects, the values for $\Gamma_{low}$ are very low and unphysical; hence, excess absorption appears to be a better explanation for these objects.

Although neither PKS~0537$-$286 nor 87GB~150844.6+571424 required excess absorption, both spectra are noticeably better fitted with a broken, rather than a single, power-law. However, these continua curve in the opposite direction from the (possibly) absorbed sources, becoming steeper at lower energies.

\begin{table*}
\begin{center}
\caption{Broken power-law fits to the spectra which show evidence for excess absorption; in these fits only Galactic N$_{H}$ has been included. $\Gamma_{\rm low}$/$\Gamma_{\rm high}$ signify the power-law slope fitting the low/high energy end of the spectrum respectively. The break energy is given in the rest frame of the object. The penultimate column restates the $\chi^{2}$ value obtained from a power-law plus excess rest-frame absorption model, first given in Table~\ref{fits}, except for the two marked $^{*}$, for which no absorption was statistically required. The final column gives the change in $\chi^{2}$ between the fits. Here a positive value indicates that the $\chi^{2}$ value is higher for the broken power-law compared to the absorption, meaning the fit is worse.}
\label{curv}
\begin{tabular}{p{3.0truecm}p{2.0truecm}p{2.0truecm}p{2.0truecm}p{1.8truecm}p{2.5truecm}p{1.0truecm}}
\hline
QSO & $\Gamma_{\rm low}$ & Break energy (keV) & $\Gamma_{\rm high}$ & $\chi^{2}$/dof & N$_{\rm H}$ $\chi^{2}$/dof & $\Delta\chi^{2}$ ($\Gamma$-N$_{\rm H}$)\\
\hline
{\bf RADIO LOUD}\\
S5~0014+81  & 0.53$^{+0.19}_{-0.28}$ & 3.62$^{+0.30}_{-0.28}$ & 1.55~$\pm$~0.02 & 1279/1158 & 1261/1159 & +18\\
RBS~315 & 0.21$^{+0.06}_{-0.08}$ &3.87~$\pm$~0.11 & 1.20~$\pm$~0.01 & 1915/1736 & 1819/1373 & +96\\ 

PKS~0438$-$43 & 0.38$^{+0.29}_{-0.90}$  & 2.77$^{+0.31}_{-0.45}$  & 1.66~$\pm$~0.03  & 490/506 & 482/507 & +8\\

PMN~0525$-$3343 & 1.10$^{+0.17}_{-0.15}$ & 4.90$^{+1.23}_{-0.52}$ & 1.71$^{+0.05}_{-0.04}$ & 470/432 & 466/433 & +4\\

PKS~0537$-$286 & 1.56$^{+0.07}_{-0.06}$ & 9.19$^{+1.11}_{-1.44}$ & 1.28~$\pm$~0.04 & 1009/847 & 1048/849$^{*}$ & -39\\ 

S5~0836+71 & 0.59$^{+0.21}_{-0.55}$ & 1.62$^{+0.09}_{-0.12}$  & 1.36~$\pm$~0.01  & 2569/2220 & 2585/2221 & -16\\

RX J1028.6$-$0844 & 1.12$^{+0.09}_{-0.11}$ & 6.06$^{+0.18}_{-0.92}$ & 1.47~$\pm$~0.04 & 367/361 & 371/362 & -4\\

RX J122135.6+280613 & 0.25$^{+0.88}_{-3.16}$ & 2.50$^{+2.08}_{-0.51}$ & 1.35$^{+0.05}_{-0.04}$ & 172/195 & 173/196 & -1\\

87GB 142825.9+421804 & 0.85$^{+0.31}_{-0.63}$ & 3.71$^{+0.69}_{-0.54}$ & 1.71~$\pm$~0.03 & 364/399 & 364/400 & 0\\

87GB 150844.6+571424 & 1.62$^{+0.04}_{-0.07}$ & 17.12$^{+5.28}_{-6.66}$ & 1.08$^{+0.31}_{-0.37}$ & 115/96  &  123/98$^{*}$ & -8\\

PKS~2126$-$15 & 0.15$^{+0.23}_{-0.31}$  & 2.87$^{+0.17}_{-0.20}$  & 1.39~$\pm$~0.01  & 1402/1222 & 1378/1223 & +24\\

{\bf BAL}\\
 
APM 08279+5255 & $-$0.38$^{+0.44}_{-0.79}$ & 4.01~$\pm$~0.39 & 0.82~$\pm$~0.06 & 252/200 & 251/201 & +1\\

RX~J0911.4+0551 & 0.92$^{+0.22}_{-0.29}$ & 6.88$^{+2.13}_{-1.79}$ & 1.96$^{+0.44}_{-0.33}$ & 56/63 & 63/64 & -7\\

Q 1246-057 & 1.31$^{+0.24}_{-0.27}$ & 10.45$^{+5.42}_{-4.55}$ & 3.83$^{+17.90}_{-1.87}$ & 59/67 & 61/68 & -2\\

CSO 0755 & 0.96$^{+0.26}_{-0.23}$ & 4.09$^{+1.20}_{-0.26}$ & 1.92$^{+0.15}_{-0.09}$ & 105/142 & 112/143 & -7\\

\hline
\end{tabular}
\end{center}
\end{table*}

The existence of Compton reflection was also investigated in each QSO, after modelling the apparent absorption at lower energies; the lack of iron line emission suggests that no reflection components should be required. Only PKS~0438$-$43 seemed to reveal evidence for a flattening of the spectrum at higher energies (Table~\ref{hardfits}), with the other values being consistent with zero. Including the reflection component improved the fit statistic to $\chi^{2}$/dof~=~472/506  [F-test probability of 1.1~$\times$~10$^{-3}$, compared to the power-law plus absorption model (Fit 4 in Table~\ref{fits})]. The photon index then became $\Gamma$~=~1.89~$\pm$~0.06, with  N$_{\rm H}$~=~(1.46~$\pm$~0.13)~$\times$~10$^{22}$~cm$^{-2}$, for a reflection parameter of R~=~0.62~$\pm$~0.28. 

However, since the upper limit for iron emission in PKS~0438$-$43 is 33~eV, this precludes the presence of reflection; hence, the deviation from the simple power-law must be due to intrinsic curvature of the spectrum. The broken power-law model was found to be an equally acceptable fit to the PKS~0438$-$43 spectrum (Table~\ref{curv}).

Summarising the work in this section, 14 of the 29 QSOs can be fitted with a single power-law; of these, five are RLQs, seven RQQs and two BAL QSOs. The remaining 15 objects show more complex continua, requiring either excess absorption or a broken power-law. Nine of the 16 RLQs have intrinsic absorption above the Galactic value, with typical columns around 10$^{22}$~cm$^{-2}$. Four of the six BAL QSOs are also absorbed (with X-ray columns of a few 10$^{22}$~cm$^{-2}$, consistent with other BAL QSOs) while none of the RQQs shows evidence for excess absorption.

\section{Discussion}
\label{disc}

This paper presents {\it XMM-Newton} observations of a sample of 29 high-redshift (z$>$2), luminous (L$_{0.3-10 \rm{keV}}$~=~1~$\times$~10$^{45}$ -- 2~$\times$~10$^{48}$ erg~s$^{-1}$) radio-loud and radio-quiet QSOs. Below follows a discussion of the statistical properties of the high-z sample.

\subsection{2--10~keV}

\begin{figure}
\begin{center}
\includegraphics[clip,width=6.4cm,angle=-90]{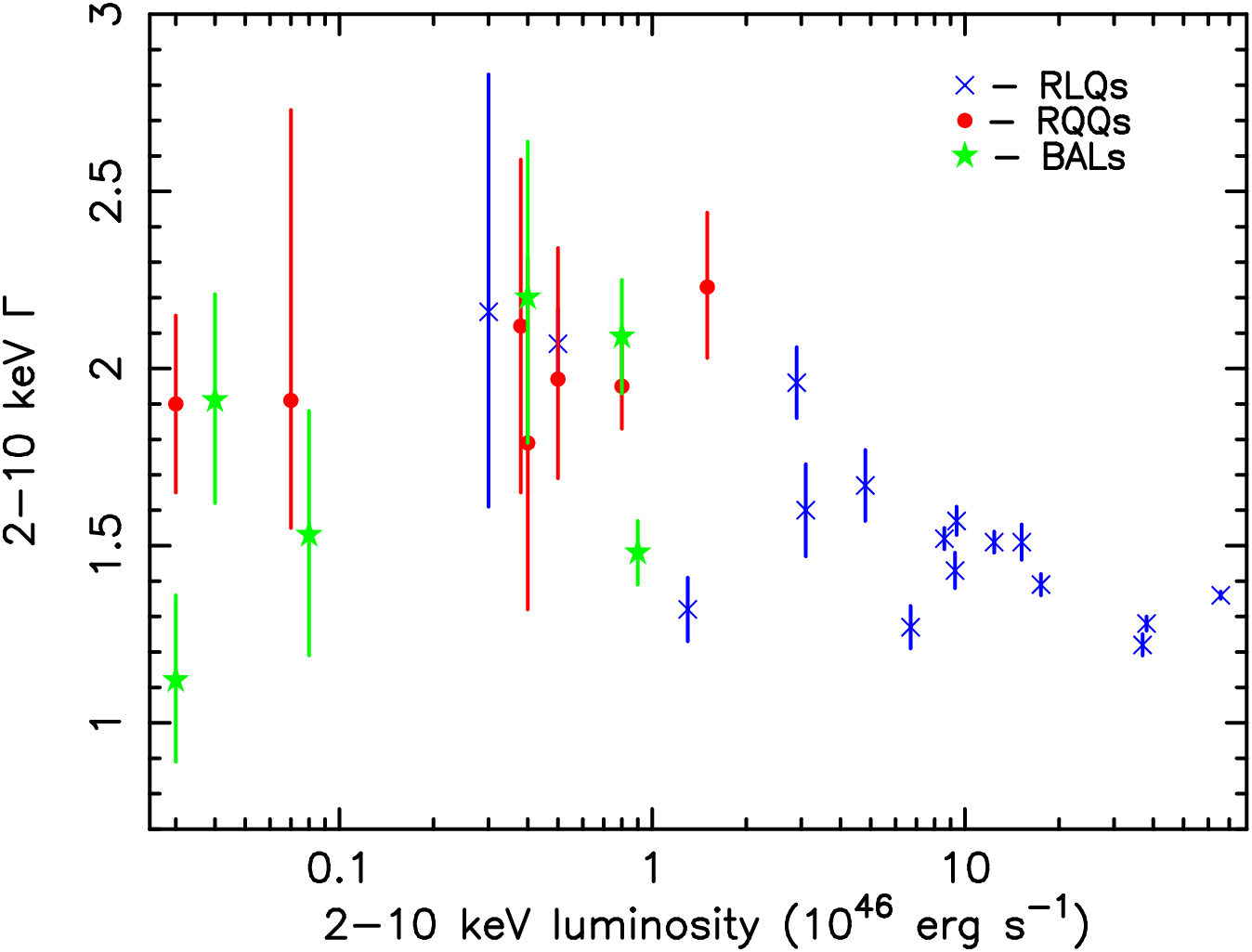}
\includegraphics[clip,width=6.4cm,angle=-90]{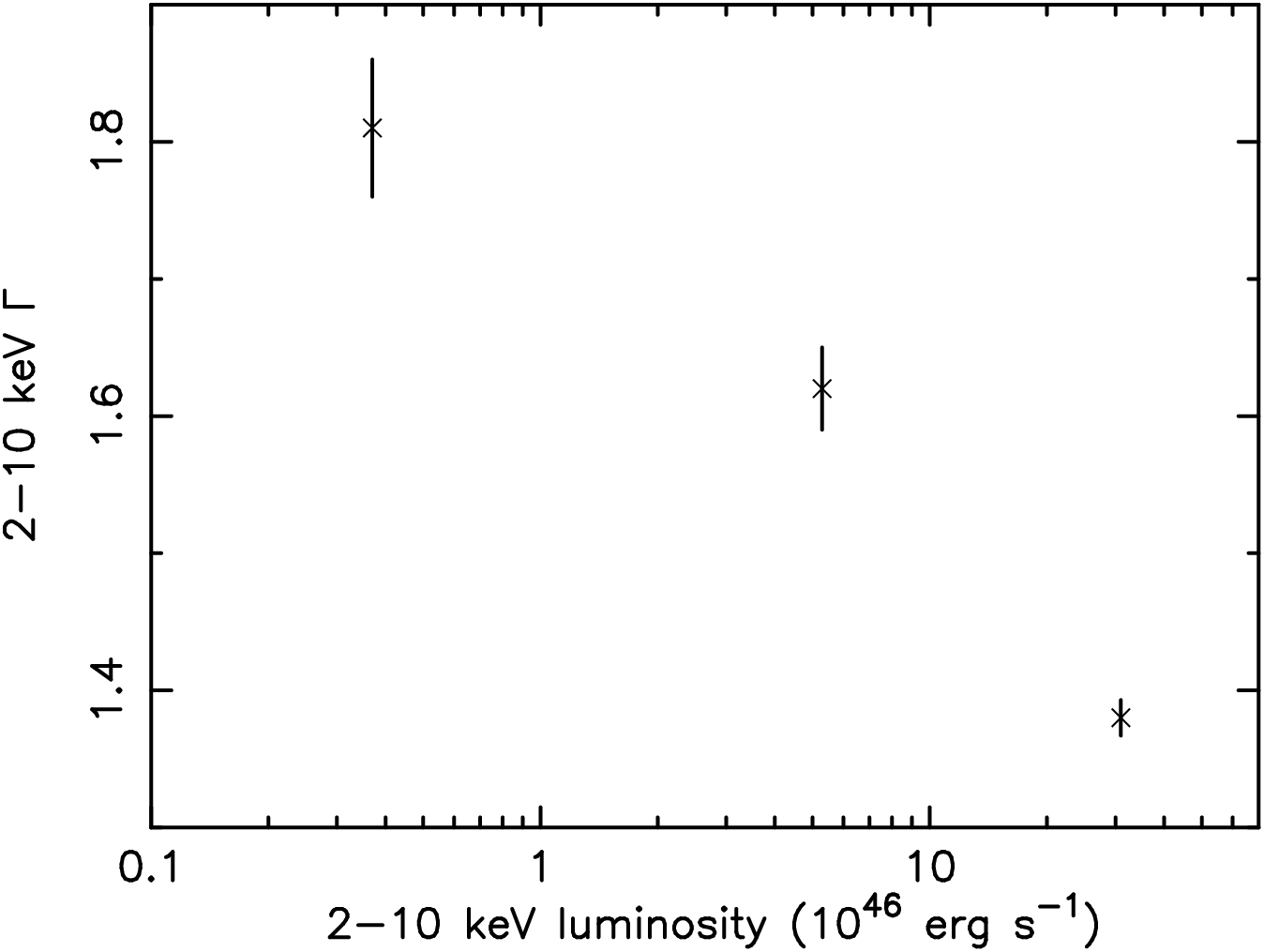}
\caption{The more luminous QSOs tend to have flatter 2-10 keV (rest-frame) photon indices. Here the measured luminosities for APM~08279+5255, RX~J0911+0551 and HE~1104+0551 have been decreased by factors of 142, 17 and 12 respectively to account for gravitational lensing effects (Dai \etal 2004). The lower plot uses three luminosity bins ($<$1~$\times$~10$^{46}$, 1-10~$\times$~10$^{46}$, $>$10~$\times$~10$^{46}$) to show this more clearly, with RLQs, RQQs and BALs all grouped together.}
\label{gamma-lum}
\end{center}
\end{figure}

Over the 2--10~keV rest-frame band, the mean power-law slopes were found to be: $\Gamma_{\rm RL}$~=~1.55~$\pm$~0.04; $\Gamma_{\rm RQ}$~=~1.98~$^{+0.17}_{-0.13}$; $\Gamma_{\rm BAL}$~=~1.80~$^{+0.10}_{-0.11}$. RLQs are known to have flatter slopes than the RQQs, with typical slopes of $\Gamma_{\rm RL}$~$\sim$~1.6--1.7, compared with $\Gamma_{\rm RQ}$~$\sim$~1.9 (Williams \etal 1992; Reeves \etal 1997; Reeves \& Turner 2000; Page \etal 2003; Piconcelli \etal 2005). The RLQs here have rather flat slopes, which could, conceivably, be an evolutionary effect. Bechtold \etal (2003) analysed a sample of high-z, radio-quiet AGN, finding that the average X-ray spectral index was flatter for the more distant objects, both due to the increased redshift and the higher luminosities. Grupe \etal (2005), however, found that $\Gamma$ was steep (2.23~$\pm$~0.48) for their high-z RQQs.
Plotting 2--10~keV $\Gamma$ against luminosity for the sample of objects in this paper agrees with the finding of Bechtold \etal (Figure~\ref{gamma-lum}), with weighted linear regression giving a slope of ($-$2.5~$\pm$~0.3)~$\times$~10$^{-3}$. In this sample, however, the correlation only exists between $\Gamma$ and luminosity, not redshift; this is consistent with there being no spectral evolution over time.

This relationship between $\Gamma$ and luminosity is, however, due to the dominance of the high-luminosity beamed, flat spectrum RLQs; considering only the RQQs and BAL QSOs in the sample, no correlation is present. The average photon index for the high-z RQQs in the current sample (1.98~$^{+0.17}_{-0.13}$) is very typical of low to medium-z AGN (Nandra \etal 1997a; George \etal 2000; Reeves \& Turner 2000; Page \etal 2003; Porquet \etal 2004; Piconcelli \etal 2005). This suggests (in agreement with earlier works, e.g., Reeves \& Turner 2000;  Page \etal 2004a; Shemmer \etal 2005; Vignali \etal 2005) that there is no evolution of the spectrum with luminosity or redshift (for radio-quiet objects).

An alternative method for obtaining a mean spectral shape is to fit all the data simultaneously, with a constant of normalisation to account for the flux differences between the different spectra. Doing this leads to a photon index of $\Gamma$~=~1.99~$\pm$~0.05 ($\chi^{2}$/dof~=~260/241) for the RQQ spectra (over 0.3--10~keV). Upper limits of EW~$<$~108~eV and R~$<$~0.53 for the equivalent width of a line at 6.4~keV and reflection parameter respectively were also derived.

The photon index for the six BAL QSOs is formally consistent with that for the normal RQQs. Comparison with other BAL QSOs published in the literature (e.g., Wang \etal 1999; Gallagher \etal 2001; Green \etal 2001; Chartas, Brandt \& Gallagher 2003; Gallagher \etal 2004) confirms that the mean slope found here is typical for such objects. This implies that the underlying X-ray continuum and central engine is the same as for RQQs, but the BAL QSO is being observed through the outflow.

Of the objects in this sample, none showed a significant Fe K$\alpha$ line. The absence of iron lines is further confirmation of the X-ray Baldwin effect (Iwasawa \& Taniguchi 1993; Nandra \etal 1997b; Page \etal 2004b), whereby the most luminous AGN, both radio-loud and radio-quiet, do not tend to reveal iron emission. In the case of the RQQs, many of the iron line equivalent widths are poorly constrained, because such AGN are less X-ray bright (for a given optical luminosity) than their radio-loud equivalents, as mentioned in the introduction.
The lack of reflection in the RLQs is not entirely surprising, since the radio-loud objects are believed to be dominated by radio-jets. This means that the Doppler-boosted jet emission can be much stronger than any reflection component, thus diluting any features. 

\subsection{Excess absorption}

13 of the 29 QSOs are better modelled by the inclusion of an additional absorption component; of these, nine (out of the 16 in the sample) are radio-loud, with four (out of six) being BALs. None of the seven radio-quiet objects require excess N$_{\rm H}$. Radio-loud objects have often previously been found to be absorbed (e.g., Elvis \etal 1994; Brinkmann \etal 1997; Reeves \& Turner 2000), while their radio-quiet counterparts are more seldom absorbed (Canizares \& White 1989; Lawson \& Turner 1997; Yuan \etal 1998; Page \etal 2003). Where Reeves \& Turner (2000) analysed the same QSOs as those presented here, the values found for N$_{\rm H}$ using the {\it ASCA} data were either in agreement with those from {\it XMM-Newton} (PKS~0438$-$43, S5~0836+71, 87GB~150844.6+571424, PKS~2000$-$330 and PKS~2126$-$15), or larger (S5~0014+81, PKS~0537$-$286, PKS~2149$-$306 and HE~1104$-$1805). RX~J1028.6$-$0844 was also found to show strong (2~$\times$~10$^{23}$ cm$^{-2}$) absorption by {\it ASCA} (Yuan \etal 2000), while the {\it XMM-Newton} data presented here, although absorbed, have a column density almost a factor of ten lower. Also, a previous observation, reported in Grupe \etal (2004), find the spectrum to be largely unabsorbed. 

It is possible that this disagreement is due to the degradation of the {\it ASCA} SIS low energy response (see Yaqoob \etal 2000). The effect of this loss of response is an underestimate of the soft X-ray flux, which can manifest itself as an increase in N$_{\rm H}$. However, since the {\it ASCA} and {\it XMM-Newton} observations are non-simultaneous, the absorbing columns could have changed. The data in Reeves \& Turner (2000) show a mean column density of $\sim$~1.3~$\times$~10$^{22}$~cm$^{-2}$, while the radio-loud quasars here give a very similar average of $\sim$~1.4~$\times$~10$^{22}$~cm$^{-2}$ for the objects with detected N$_{\rm H}$.

Agreeing with earlier work using smaller samples of objects (e.g., Elvis \etal 1994; Cappi \etal 1997; Reeves \etal 1997; Fiore \etal 1998; Reeves \& Turner 2000; Bassett \etal 2004) found that the more distant QSOs were more heavily (intrinsically) absorbed -- see their figure 9. To compare with the {\it ASCA} results, a similar plot was produced for objects observed by {\it XMM-Newton}. This includes absorbing columns for the QSOs in this paper, other datasets possessed by the authors and any other corresponding values obtained from the literature for objects at a redshift greater than 0.01. Figure~\ref{nh-z} shows the plot, while, in Appendix B,  Table~\ref{nh}, lists the objects, their redshift and the {\it neutral} absorbing column; for sources where there was evidence for a warm absorber, this was fitted before determining the additional cold column.

\begin{figure}
\begin{center}
\includegraphics[clip,width=6.4cm,angle=-90]{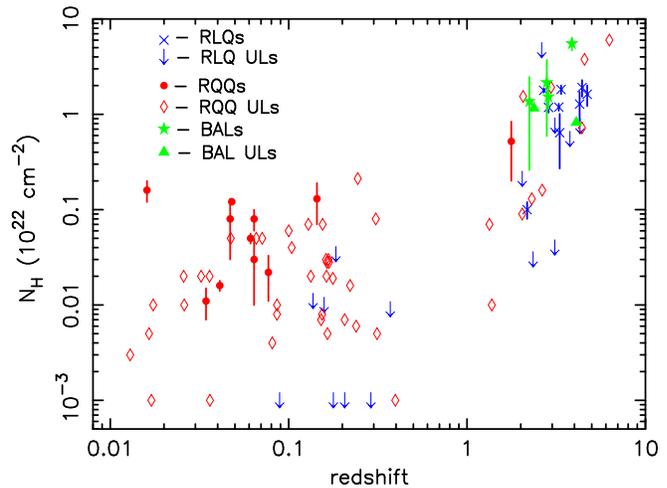}
\caption{Intrinsic column density (N$_{\rm H}$) plotted against redshift. The Galactic absorption has been separately accounted for. RLQs are plotted as crosses/arrows, RQQs as circles/diamonds and BALs as stars/triangles, for detections/upper limits (90~per~cent) respectively.}
\label{nh-z}
\end{center}
\end{figure}

Because of the presence of censored (upper limit) data, the {\sc asurv} software (Astronomy Survival Statistics; Feigelson \& Nelson 1985; Isobe, Feigelson \& Nelson 1986) was used to determine the line of best fit through the data points. Although this method takes into account that some of the measurements are upper limits, the errors on the actual values are not accounted for. The outcome of this analysis was that there is a positive correlation between redshift and intrinsic N$_{\rm H}$ (linear regression slope of 0.61~$\pm$~0.10); that is, the higher-redshift QSOs are more strongly absorbed. Kendall's Tau (also within {\sc asurv}) gave a probability of $>$99.9~per~cent for the correlation. Spearman's Rank gave a probability of $>$99.99\% for a positive correlation between the intrinsic N$_{\rm H}$ and redshift, both taking the upper limit N$_{\rm H}$ values as detections and also setting them as the mid-points of the interval. The actual detections alone also showed this strong, positive correlation.

The separate groups of RQQs and RLQs were also investigated, to determine whether there is a difference between the intrinsic absorption in radio-quiet and radio-loud objects. In Page \etal (2003) the authors compared a sample of {\it XMM-Newton} serendiptitous radio-quiet AGN with {\it ASCA}-detected radio-loud objects, concluding that there was a difference between the absorption in these groups. The QSOs in that analysis were lower-redshift than those presented here, at z~$<$~3. Using the two-sample tests within {\sc asurv} for the current sample gave a probability of 85-92~per~cent for the RLQs and RQQs having different absorption, with N$_{\rm H}$ being higher for the RLQs. Although this difference is marginal, it should be noted that none of the RQQs in this sample shows any evidence for an excess column, while more than half of the RLQs are absorbed. These results are not conclusive, though they do tend to indicate that there is a difference in the intrinsic absorbing columns found in radio-loud and radio-quiet AGN.
Recently, however, Grupe \etal (2005) found that their high-redshift RLQs were not more highly absorbed than the radio-quiet counterparts; only some of their targets overlap with the sample prsented here. Clearly work remains to be done in this area, with more good quality observations of high-z quasars, both radio-loud and radio-quiet, required.

It should be noted that, since RLQs can, in general, be studied out to higher redshifts (see the introduction), the N$_{\rm H}$-z correlation and the fact that RLQs may be more strongly absorbed than RQQs could be due to the same effect. The Partial Spearman Rank test can be used to try and determine whether, for example, the correlation between N$_{\rm H}$ and R$_{\rm L}$ is simply due to high-redshift quasars frequently being radio-louder than the nearby ones; in this case, the `real' correlation would be N$_{\rm H}$-z. Unfortunately, because many of the estimates for the absorbing column are upper limits, as are the R$_{\rm L}$ values for the RQQs, it was not possible to obtain a definitive answer on this subject. 
Taking the upper limits as actual measurements gave similar probabilities for the N$_{\rm H}$-z and N$_{\rm H}$-R$_{\rm L}$ correlations, while halving each of the upper limits gave a slightly higher probability that N$_{\rm H}$ is correlated with redshift only due to its correlation with R$_{\rm L}$. In each case, however, individual Spearman Rank tests for N$_{\rm H}$-z and  N$_{\rm H}$-R$_{\rm L}$ implied that N$_{\rm H}$ was more strongly correlated with redshift.



\subsubsection{Damped Lyman-$\alpha$ systems}

It is possible that some of the neutral absorption identified in these spectra is not intrinsic to the QSOs, but, rather, is due to intervening damped Lyman-$\alpha$ (DLA) systems. DLAs are high density (N$_{\rm HI} > 2~\times~10^{20}$ atoms cm$^{-2}$) systems, which contain most of the neutral gas found at high redshifts; the Lyman-$\alpha$ forest lines are far more numerous, but only make up a small fraction of the gas. Searching the literature, it was found that some of the QSOs in the current sample do have intervening DLAs. However, these do not tend to be the objects which show the highest excess absorption (see Table~\ref{dla}). Hence, most of the absorption in these spectra is likely to be intrinsic to the QSOs themselves.

\begin{table}
\begin{center}
\caption{The QSOs for which intervening Damped Lyman-$\alpha$ systems have been firmly identified or discounted. The reference for the information is given in the final column.} 
\label{dla}
\begin{tabular}{p{2.2truecm}p{0.9truecm}p{1.8truecm}p{2.5truecm}}
\hline
Object & DLA ? & {\it XMM} N$_{\rm H}$ & DLA reference\\
 & &(10$^{22}$ cm$^{-2}$)\\
\hline

S5~0014+81 & no & 1.82~$\pm$~0.19 & Bechtold \etal (1994)\\
Q~0420$-$388 & yes & $<$0.76 & Elvis \etal (1994)\\
PMN~0525$-$3343 & no & 1.91~$\pm$~0.38 & P{\' e}roux \etal (2001)\\
PKS~0537$-$286 & yes &$<$0.04 & Kanekar \& Chengalur (2003)\\
RX~J1028.6$-$0844 & yes & $<$1.17 & P{\' e}roux \etal (2001)\\
Q~0000$-$263 & yes & $<$0.82 & Lu \etal (1996)\\
LBQS~0053$-$0134 & no & $<$1.54 & Storrie-Lombardi \& Wolfe (2000)\\
HE~1104$-$1805 & yes & $<$0.13 & Lanfranchi \& Fria{\c c}a (2003)\\
BR~2237$-$0607 & yes & $<$3.77 & Lu \etal (1996)\\

\hline
\end{tabular}
\end{center}
\end{table}

\subsubsection{Origins for the absorbing column}

Assuming that the absorption is intrinsic to the objects, there are various possible origins for the absorbing material (see, e.g., Elvis \etal 1994; Cappi \etal 1997). The quasars themselves could be surrounded by a region of dust or gas (Rieke, Lebofsky \& Kinman 1979; Sanders \etal 1989; Webster \etal 1995), which would lead to obscuration of the AGN. Alternatively, since some RLQs can be located in clusters of galaxies (Yee \& Ellingson 1993), the absorbing material measured may be due to a cooling flow (White \etal 1991; Elvis \etal 1994). Cold clouds of material close into the central regions of the AGN may exist (Guilbert \& Rees 1988; Celotti, Fabian \& Rees 1992), which could lead to both reflection features and absorption. That being said, no Compton reflection humps or iron lines are found in the present sample.

None of the absorbed (non-BAL) quasars have significantly reddened optical spectra, indicating that the dust/gas ratio is low. The absorbing material should, therefore, be located close to the central engine, in order for the dust to sublimate. 

The possibility exists that the material is related to the warm absorbers seen in Seyfert 1s; the column densities of $\sim$~10$^{22}$~cm$^{-2}$ are similar to a number of these, although the level of ionisation cannot be tied down. 
Because of the bandpass covered for these high-z quasars, it may be just the continuum curvature below the iron K band which is being observed. Examples of this (convex) spectral curvature are seen in NGC~3783 (Kaspi \etal 2002; Reeves \etal 2004) and NGC~3516 (Turner \etal 2005). In order to constrain such an absorber, lines and edges would need to be measured at high resolution and signal-to-noise, which will not be possible for these sources until the next generation of instruments (e.g., {\it Con-X} and {\it XEUS}).

If RLQs do tend to be more absorbed than RQQs, then it is possible that the absorbing material is somehow linked to the radio jet - perhaps some kind of confining medium, which helps to collimate the jets (e.g., Crawford \& Vanderriest 1997; Ferrari 1998).


\section{Summary}


As has been found in previous samples (Williams \etal 1992; Reeves \etal 1997; Reeves \& Turner 2000; Page \etal 2003; Piconcelli \etal 2005), RLQs have
flatter spectral indices than RQQs, suggesting the presence of an
extra component due to a relativistic jet. The lack of any iron emission
lines or a reflection component in the RLQs is also consistent with this
interpretation. However none of the radio-quiet quasars shows significant
Fe emission either, but this is probably due to the poorer photon
statistics in these objects.

There is no evidence for any evolution of the X-ray continuum for
radio-quiet quasars with either redshift or luminosity. The mean photon
index for this high z, high luminosity sample is consistent with those
in lower z or lower luminosity samples (Nandra \etal 1997a; George \etal 2000; Reeves \& Turner 2000; Page \etal 2003; Porquet \etal 2004; Piconcelli \etal 2005). Thus the X-ray properties of the quasars do not appear to
evolve with redshift or luminosity and may just simply scale with black hole mass
and luminosity.

The continuum properties of six high-z BAL QSOs have also been measured in this paper. Four of
these six objects show large absorbing columns ($>$10$^{22}$~cm$^{-2}$ in the rest frame),
presumably due to intersecting the outflowing material along the line of sight.
After correcting for absorption, the continuum photon indices of the BAL
QSOs are the same as the non-BAL RQQs, consistent with the difference
solely being due to the inclination of the observer with respect to the
outflow.

Over half of the RLQs in this sample (nine out of 16) show evidence for
intrinsic absorption in excess of the Galactic value. In the QSO rest
frame the absorption is typically 10$^{22}$ cm$^{-2}$, in agreement with earlier
studies with {\it ROSAT}, {\it ASCA} and {\it Chandra} (Elvis \etal 1994; Brinkmann \etal 1997; Reeves \& Turner 2000; Bassett \etal 2004). There is no correlation
between the absorption and the occurence of any dampled Ly$\alpha$ system along
the line of sight, suggesting that the absorbers are associated with the quasar or
its host galaxy. The lack of reddening in the optical spectra of the
quasars favours the absorber being located close (within a few parsec) of
the quasar central engine. Absorption is not found in low-z radio-loud
quasars [e.g., 3C 273 (Page \etal 2004d);  B2~1028+31, B2~1128+31, B2~1721+34 (Page \etal 2004c)], consistent with the N$_{H}$-z correlation found
from {\it ASCA} quasar samples (e.g., Reeves \& Turner 2000). High throughput and
high resolution X-ray spectroscopy will be needed to determine the origin
of the low energy absorption, through the detection of
absorption lines/edges in the quasar spectra.

\section{ACKNOWLEDGMENTS}
The work in this paper is based on observations with {\it
XMM-Newton}, an ESA
science mission, with instruments and contributions directly funded by
ESA and NASA. The authors would like to thank the EPIC Consortium for all their work during the calibration phase, 
and the SOC and SSC teams for making the observation and analysis
possible. We also thank the referee, Delphine Porquet, for useful comments and suggestions.
This research has made use of the NASA/IPAC Extragalactic
Database (NED), which is operated by the Jet Propulsion Laboratory,
California Institute of Technology, under contract with the National
Aeronautics and Space Administration.

\appendix
\section{Previous Work}

The data for some of these {\it XMM-Newton} observations has been previously published. As already mentioned, PKS~0537$-$286 was analysed in Reeves \etal (2001), with the current results being in agreement. 

Worsley \etal (2004a) published a paper about the absorption in PMN~0525$-$3343, using data from eight separate observations. Their value for the intrinsic N$_{\rm H}$ (for a joint fit to all the datasets, rather than the single observation considered in this paper) is consistent with the results here at the 90~per~cent level. Although there are no clear indications of a warm absorber, they conclude that cold absorption is inconsistent with the optical data and that an ionised absorber would be more likely. Both of the {\it XMM-Newton} observations for 87GB~142825.9+421804 were also analysed by Worsley \etal (2004b), while only the second (longer) dataset is used here. Once again, they favour a warm absorber model, although intrinsic cold absorption is not ruled out. The column density for cold absorption is in agreement with the value presented in this paper.

Ferrero \& Brinkmann (2003) analysed the spectra of PKS~2126$-$15, PKS~2149$-$306, Q~0000$-$263 and Q~1442+2931. The evidence for absorption in PKS~2149$-$306 is marginal and only the Galactic column is required for Q~0000$-$263 and Q~1442+2931; these results are in agreement with the current work. PKS~2126$-$15 is found to be absorbed, with neutral N$_{\rm H}$~$\sim$~1.4~$\times$~10$^{22}$ cm$^{-2}$; this is slightly larger than the value found here, but their spectrum was fitted down to 0.2~keV, rather than 0.3~keV.

Brocksopp \etal (2004) presented the spectra of PKS~0438$-$436, RX~J122135.6+280613, PKS~2000$-$330 and SBS~1107+487, finding that the first two required cold absorption components (with values consistent with those given here), while the other two were not significantly absorbed (also found here).

Yuan \etal (2004) analysed the observation of RX~J1028.6$-$0844 discussed in the present paper. An earlier observation (Obs. ID: 0093160701; date of observation: 2002-05-15) was covered as well and is also the subject of the paper by Grupe \etal (2004). This previous observation was much shorter ($<$5~ks in each of the EPIC instruments), so the data are not included in this paper directly; however, they were analysed and the results were in agreement with Grupe \etal (2004), in that no intrinsic absorption was found. In the more recent observation, however, there is evidence for excess N$_{\rm H}$, as also found by Yuan \etal (2004). They considered the MOS and PN cameras separately, finding that the values of N$_{\rm H}$ from the PN detector were lower than those from MOS -- possibly underestimated due to low-energy calibration uncertainties. 

Hasinger, Schartel \& Komossa (2002) discovered the same ionised Fe~K edge in the {\it XMM-Newton} spectrum of APM~08279+5255 as was found in this present analysis. 

\section{Neutral absorption values}

\begin{table}
\begin{center}
\caption{The data for Figure~\ref{nh-z}, listing the corresponding redshifts and neutral N$_{\rm H}$ values; the objects are listed by decreasing redshift. $^{*}$ marks the objects which also have a warm absorber. All values, except where the quasars are marked $^\dag$, were inferred from the authors' own work.
Values from the literature: $^{a}$ Porquet \etal (2004); $^{c}$ Gallo \etal (2004b); $^{d}$ Gallo \etal (2004a); $^{e}$ Vaughan \etal (2004).} 
\label{nh}
\begin{tabular}{p{2.8truecm}p{1.5truecm}p{2.5truecm}}
\hline
Object & redshift & N$_{\rm H}$ (10$^{22}$ cm$^{-2}$)\\
\hline
{\bf RADIO-LOUD}\\
87GB 142825.9+421804  & 4.715 & 1.62~$\pm$~0.40\\
PMN~0525$-$3343 & 4.413 & 1.91~$\pm$~0.38\\
87GB~150844.6+571424 &  4.301 & $<$0.75\\
RX~J1028.6$-$0844 & 4.276 & 1.28~$\pm$~0.44\\
PKS~2000$-$330 & 3.773 & $<$0.55\\
S5~0014+81 & 3.366 & 1.82~$\pm$~0.19\\
RX~J122135.6+280613 & 3.305 & 0.64~$\pm$~0.37\\
PKS~2126$-$15 & 3.268 & 1.19~$\pm$~0.10\\
Q~0420$-$388 & 3.11 & $<$0.76\\
PKS~0537$-$286 & 3.104 & $<$0.04\\
PKS~0438$-$43 & 2.863 & 1.18~$\pm$~0.14\\
RBS~315 & 2.69 & 1.78$^{+0.08}_{-0.09}$\\
HS~0848+1119 & 2.62 & $<$4.70\\
PKS~2149$-$306 & 2.345 &$<$0.03\\
S5~0836+71 & 2.172 & 0.10~$\pm$~0.02 \\
SBS~1345+584 & 2.039 & $<$0.21\\
PG~1512+370 $^{\dag,a}$ & 0.371 & $<$0.009\\
B2~1128+31 & 0.289 & $<$0.001\\
B2~1721+34 & 0.206 & $<$0.001\\
PG~1309+355$^{\dag,a}$ & 0.184 & $<$0.034\\
B2~1028+31 & 0.178 & $<$0.001\\
3C~273 & 0.158 & $<$0.01\\
PKS~0558$-$504 & 0.137 & $<$0.011\\
3 Zw 2 & 0.089 & $<$0.001\\

{\bf RADIO-QUIET}\\

SDSS~1030+0524 & 6.28 & $<$6.01\\
BR~2237$-$0607 & 4.558 & $<$3.77\\
PSS~J0248+1802 & 4.411 & $<$0.72\\
SBS~1107+487 & 2.958 & $<$1.92\\
Q~1442+2931 & 2.638 & $<$0.16\\
HE~1104$-$1805 & 2.305 & $<$0.13\\
LBQS~0053$-$0134 & 2.062 & $<$1.54\\
PG~1247+267 & 2.038 & $<$0.09\\
PB~5062 & 1.77 & 0.52~$\pm$~0.32\\
SBS~0909+532 & 1.376 & $<$0.01\\
PG~1634+706 & 1.334 & $<$0.07\\
PHL~1092 $^{\dag,d}$ & 0.396 & $<$0.001\\
Mrk~493 & 0.313 & $<$0.005\\
UM~269 & 0.308 & $<$0.08\\
Q~0144$-$3938$^{*}$ & 0.244 & $<$0.212\\
PG~0953+414$^{\dag,a}$ & 0.2389 & $<$0.006\\
PG~1427+480$^{\dag,a}$ & 0.221 & $<$0.016\\
PG~0947+396$^{\dag,a}$ & 0.206 & $<$0.007\\
PG~1116+215$^{\dag,a}$ & 0.177 & $<$0.019\\
PG~1322+659$^{\dag,a}$ & 0.168 & $<$0.028\\
PG~1048+342& 0.167 & $<$0.03\\
PG~1202+281$^{\dag,a}$ & 0.165 & $<$0.005\\
PG~1402+261$^{\dag,a}$ & 0.164 & $<$0.028\\
Mrk~1014 & 0.163 & $<$0.02\\
PG~1307+085$^{\dag,a}$ & 0.155 & $<$0.07\\
PG~1115+407$^{\dag,a}$ & 0.154 & $<$0.008\\



\hline
\end{tabular}
\end{center}
\end{table}
\addtocounter{table}{-1}
\begin{table}
\begin{center}
\caption{-- {\bf continued}} 
\begin{tabular}{p{2.5truecm}p{1.5truecm}p{2.5truecm}}
\hline
Object & redshift & N$_{\rm H}$ (10$^{22}$ cm$^{-2}$)\\
\hline

PG~1352+183$^{\dag,a}$ & 0.152 & $<$0.007\\
PG~1114+445$^{*}$ & 0.144 & 0.13~$\pm$~0.06\\
PG~1626+554 & 0.133 & $<$0.02\\
Mrk~876 & 0.129 & $<$0.07\\
1H~0419$-$577 & 0.104 & $<$0.04\\
PG~0804+761 & 0.100 & $<$0.06\\
HE~1029$-$1401 & 0.086 & $<$0.008\\
Mrk~1383 & 0.086 & $<$0.01\\
PG~1211+143 & 0.0809 & $<$0.004\\
Mrk~478$^{\dag,a}$ & 0.077 & 0.022~$\pm$~0.011\\ 
Mrk~205 & 0.071 & $<$0.05\\
Mrk~304$^{*}$ & 0.066 & $<$0.05\\
MR~2251$-$178 & 0.064 & 0.08~$\pm$~0.02\\
PG~0844+349 & 0.064 & 0.03~$\pm$~0.02\\
1Zw1$^{\dag,a}$ & 0.0611 & 0.05~$\pm$~0.006\\
PG~1244+026$^{\dag,a}$ & 0.048 & 0.121~$\pm$~0.001\\
Fairall 9 & 0.047 & 0.08~$\pm$~0.05\\
MCG~$-$2$-$58$-$22 & 0.047 & $<$0.05\\
1H~0707$-$495$^{\dag,d}$ & 0.0411 & 0.16~$\pm$~0.002\\
ESO~141$-$G55 & 0.036 & $<$0.02\\
Mrk~841 & 0.036 & $<$0.001\\
Mrk~509 & 0.0344 & 0.011~$\pm$~0.004\\
Ark~120 $^{\dag,e}$ & 0.0323 & $<$0.02\\
Mrk~335 & 0.026 & $<$0.02\\
Mrk~896 & 0.026 & $<$0.01\\
Mrk~359 & 0.017 & $<$0.01\\
Mrk~1044 & 0.017 & $<$0.005\\
NGC~5548 & 0.017 & $<$0.001\\
IC~4329a$^{*}$ & 0.016 & 0.16~$\pm$~0.04 \\
Mrk~766$^{*}$ & 0.0129 & 0.011~$\pm$~0.003\\


{\bf BAL}\\

Q~0000$-$263 & 4.098 & $<$0.82\\ 
APM~08279+5255 & 3.87 & 5.50$^{+0.83}_{-0.76}$\\ 
CSO 0755 & 2.86 & 1.52~$\pm$~0.38\\
RX~J0911.4+0551 & 2.80 & 2.15$^{+1.56}_{-1.19}$\\
SBS~1542+541 & 2.371 & $<$1.16\\
Q~1246$-$057 & 2.236 & 1.36$^{+1.33}_{-0.92}$\\

\hline
\end{tabular}
\end{center}
\end{table}


\begin{thebibliography}{}
\bibitem[Bassett 2004]{bas04}Bassett L.C., Brandt W.N., Schneider D.P., Vignali C., Chartas G., Garmire G.P., 2004, AJ, 128, 523
\bibitem[Bechtold 1994]{be94}Bechtold J. \etal, 1994, AJ, 108, 374
\bibitem[Bechtold 2003]{be03}Bechtold J. \etal, 2003, ApJ, 588, 119
\bibitem[Brinkmann 1997]{br97}Brinkmann W., Yuan W., Siebert J., 1997, A\&A, 319, 413
\bibitem[Brocksopp 2004]{br04}Brocksopp C., Puchnarewicz E.M., Mason K.O., C{\' o}rdova F.A., Priedhorsky W.C., 2004, MNRAS, 349, 687
\bibitem[Canizares 1989]{ca89}Canizares C.R., White J.L., 1989, ApJ, 339, 27
\bibitem[Canizares 1984]{ca84}Canizares C.R., Kruper J., 1984, 278, L99
\bibitem[Cappi 1997]{ca97}Cappi M., Matsuoka M., Comastri A., Brinkmann W., Elvis M., Palumbo G.G.C., Vignali C., 1997, ApJ, 478, 492
\bibitem[Celotti 1992]{ce92}Celotti A., Fabian A.C., Rees M.J., 1992, MNRAS, 255, 419
\bibitem[Chartas 2003]{ch03}Chartas G., Brandt W.N., Gallagher S.C., 2003, ApJ, 595, 85
\bibitem[Condon 1998]{co98}Condon J.J., Cotton W.D., Greisen E.W., Yin Q.F., Perley R.A., Taylor G.B., Broderick J.J., 1998, AJ, 115, 1693
\bibitem[Crawford 1997]{cr97}Crawford C.S., Vanderriest C., 1997, MNRAS, 285, 580 
\bibitem[Dai 2004]{dai04}Dai X., Chartas G., Eracleous M., Garmire G.P., 2004, ApJ, 605, 45
\bibitem[Dickey 1990]{di90}Dickey J.M., Lockman F.J., 1990, ARA\&A, 28, 215
\bibitem[Elvis 1994]{el94}Elvis M., Fiore F., Wilkes B., McDowell J., Bechtold J., 1994, ApJ, 422, 60
\bibitem[Fabian 1979]{fa79}Fabian A.C., 1979, Royal Soc. (London) Proc. Series A, 366, 449
\bibitem[Fabian 2001]{fa01}Fabian A.C., Celotti A., Iwasawa K., McMahan R.G., Carilli C.L., Brandt W.N., Ghisellini G., Hook I.M., 2001, MNRAS, 323, 373
\bibitem[Farrah 2004]{fa04}Farrah D., Priddey R., Wilman R., Haehnelt M., McMahon R., 2004, ApJ, 611, L13
\bibitem[Feigelson 1985]{fe85}Feigelson E.D., Nelson P.I., 1985, ApJ, 293, 192
\bibitem[Ferrari 1998]{fe98}Ferrari A., 1998, ARA\&A, 36, 539
\bibitem[Ferrero 2003]{fe03}Ferrero E., Brinkmann W., 2003, A\&A, 402, 465
\bibitem[Fiore 1998]{fi98}Fiore F., Elvis M., Giommi P., Padovani P., 1998, ApJ, 492, 79
\bibitem[Gallagher 2004]{gal04}Gallagher S.C., Brandt W.N., Wills B.J., Charlton J.C., Chartas G., Laor A., 2004, ApJ, 603, 425
\bibitem[Gallagher 2001]{gal01}Gallagher S.C., Brandt W.N., Laor A., Elvis M., Mathur S., Wills B.J., Iyomoto N., 2001, ApJ, 546, 795
\bibitem[Gallo 2004a]{ga04b}Gallo L.C., Boller Th., Brandt W.N., Fabian A.C., Grupe D., 2004a, MNRAS, 352, 744
\bibitem[Gallo 2004b]{ga04c}Gallo L.C., Tanaka Y., Boller Th., Fabian A.C., Vaughan S., Brandt W.N., 2004b, MNRAS, 353, 1064
\bibitem[George 2000]{ge00}George I.M., Turner T.J., Yaqoob T., Netzer H., Laor A., Mushotzky R.F., Nandra K., Takahashi T., 2000, ApJ, 531, 52
\bibitem[Green 2001]{gr01} Green P.J., Aldcroft T.L., Mathur S., Wilkes B.J., Elvis M., 2001, ApJ, 558, 109
\bibitem[Gregory 1991]{gr91}Gregory P.C., Condon J.J., 1991, ApJS, 75, 1011
\bibitem[Gregory 1996]{gr96}Gregory P.C., Scott W.K., Douglas K., Condon J.J., 1996, ApJS, 103, 427
\bibitem[Grupe 2005]{gr05}Grupe D., Mathur S., Wilkes B., Osmer P., 2005, submitted to AJ (astro-ph/0501521)
\bibitem[Grupe 2004]{gr04}Grupe D., Mathur S., Wilkes B., Elvis M., 2004, AJ, 127, 1
\bibitem[Grupe 2004]{gr03}Grupe D., Mathur S., Elvis M.,2003, AJ, 126, 1159 
\bibitem[Guilbert 1988]{gu88}Guilbert P.W., Rees M.J., 1988, MNRAS, 233, 475
\bibitem[Hasinger 2002]{ha02}Hasinger G., Schartel N., Komossa S., 2002, ApJ, 573, L77
\bibitem[Holt 2004]{ho04}Holt J., Benn C.R., Vigotti M., Pedani M., Carballo R., Gonz{\' a}lez-Serrano J.I., Mack K.-H., Garcia B., 2004, MNRAS, 348, 847
\bibitem[Isobe 1986]{is86}Isobe T., Feigelson E.D., Nelson P.I., 1986, ApJ, 306, 490
\bibitem[Iwasawa 1993]{iw93}Iwasawa K., Taniguchi Y., 1993, ApJ, 413, L15
\bibitem[Kanekar 2003]{ka03}Kanekar N., Changalur J.N., 2003, A\&A, 399, 857
\bibitem[Kaspi 2000]{kas00}Kaspi S., Brandt W.N., Schneider D.P., 2000, ApJ, 119, 2031
\bibitem[Kaspi 2002]{kas02}Kaspi S. \etal, 2002, ApJ, 574, 643
\bibitem[Kellerman 1989]{ke89}Kellerman K.I., Sramek R., Schmidt M., Schaffer D.B., Green R., 1989, AJ, 98, 1995
\bibitem[Kuhn 2001]{ku01}Kuhn O., Elvis M., Bechtold J., Elston R., 2001, ApJS, 136, 225
\bibitem[Lanfranchi 2003]{la03}Lanfranchi G.A., Fria{\c c}a A.C.S., 2003, MNRAS, 343, 481
\bibitem[Lawson 1997]{la97}Lawson A., Turner M.J.L., 1997, MNRAS, 288, 920
\bibitem[Lonsdale 1993]{lo93}Lonsdale C.J., Barthel P.D., Miley G.K., 1993, ApJS, 87, 63
\bibitem[Lu 1996]{lu96}Lu L., Sargent W.L.W., Barlow T.A., Churchill C.W., Vogt S.S., 1996, ApJS, 107, 475
\bibitem[Madejeski 1991]{ma91}Madejski G.M., Mushotzky R.F., Weaver K.A., Arnaud K.A., Urry C.M., 1991, ApJ, 370, 198
\bibitem[Nandra 1997]{na97a}Nandra K., George I.M., Mushotzky R.F., Turner T.J., Yaqoob T., 1997a, ApJ, 477, 602
\bibitem[Nandra 1997]{na97b}Nandra K., George I.M., Mushotzky R.F., Turner T.J., Yaqoob T., 1997b, ApJ, 488, L91
\bibitem[Page 2004]{pa04a}Page K.L., Reeves J.N., O'Brien P.T., Turner M.J.L., Worrall D.M., 2004a, MNRAS, 353, 133
\bibitem[Page 2004]{pa04b}Page K.L., O'Brien P.T., Reeves J.N., Turner M.J.L., 2004b, MNRAS, 347, 316
\bibitem[Page 2004]{pa04c}Page K.L., Schartel N., Turner M.J.L., O'Brien P.T., 2004c, MNRAS, 352, 523 
\bibitem[Page 2004]{pa04d}Page K.L., Turner M.J.L., Done C., O'Brien P.T., Reeves J.N., Sembay S., Stuhlinger M., 2004d, MNRAS, 349, 57
\bibitem[Page 2003]{pa03}Page K.L., Turner M.J.L., Reeves J.N., O'Brien P.T., Sembay S., 2003, MNRAS, 338, 1004
\bibitem[Peroux 2001]{pe01}P{\' e}roux C., Storrie-Lombardi L.J., McMahon R.G., Irwin M., Hook I.M., 2001, AJ, 121, 1799
\bibitem[Piconcelli 2005]{pic05}Piconcelli E., Jimenez-Bail{\'o}n E., Guainazzi M., Schartel N., Rodr{\'\i}guez-Pascual P.M., Santos-Lle{\'o} M., 2005, A\&A, 432, 15
\bibitem[Porquet 2004]{por04}Porquet D., Reeves J.N., O'Brien P.T., Brinkmann W., 2004, A\&A, 422, 85
\bibitem[Reeves 2004]{re04}Reeves J.N., Nandra K., George I.M., Pounds K.A., Turner T.J., Yaqoob T., 2004, ApJ, 602, 648
\bibitem[Reeves 2001]{re01}Reeves J.N. \etal, 2001, A\&A, 365, L116
\bibitem[Reeves 2000]{re00}Reeves J.N., Turner M.J.L., 2000, MNRAS, 316, 234
\bibitem[Reeves 1997]{re97}Reeves J.N., Turner M.J.L., Ohashi T., Kii, T., 1997, MNRAS, 292, 468
\bibitem[Rieke 1979]{ri79}Rieke G.H., Lebofsky M.J., Kinman T.D., 1979, ApJ, 232, L151
\bibitem[Sanders 1989]{sa89}Sanders D.B. Phinney E.S., Neugebauer G., Soifer B.T., Matthews K., 1989, ApJ, 347, 29
\bibitem[Schartel 1996]{sc96}Schartel N., Walter R., Fink H.H., Tr{\" u}mper J., 1996, A\&A, 307, 33
\bibitem[Shemmer 2005]{sh05}Shemmer O., Brandt W.N., Vignali C., Schneider D.P., Fan X., Richards G.T., Strauss M.A., 2005, ApJ, in press (astro-ph/0505482)
\bibitem[Storrie-Lombardi 2000]{st00}Storrie-Lombardi L.J.  Wolfe A.M., 2000, ApJ, 543, 552
\bibitem[Struder 2001]{st01}Str{\" u}der L. \etal, 2001, A\&A, 365, L18
\bibitem[Thorne 1974]{th74}Thorne K.S., 1974, ApJ, 191, 507
\bibitem[Turner 2005]{tu05}Turner T.J., Kraemer, S.B., George, I.M., Reeves J.N., Bottorff M.C., 2005, ApJ, 618, 155
\bibitem[Turner 1989]{tu89}Turner T.J., Pounds K.A., 1989, MNRAS, 240, 833
\bibitem[Turner 2001]{tu01}Turner M.J.L. \etal, 2001, A\&A, 365, L27
\bibitem[Turnshek 1991]{turn91}Turnshek D.A., Macchetto F., Bencke M.V., Hazard C., Sparks W.B., McMahan R.G., 1991, ApJ, 382, 26
\bibitem[Vaughan 2004]{va04}Vaughan S., Fabian A.C., Ballantyne D.R., De Rosa A., Piro L., Matt G., 2004, MNRAS, 351, 193
\bibitem[Vignali 2005]{vi05}Vignali C., Brandt W.N., Schneider D.P., Kaspi S., 2005, AJ, 129, 2519
\bibitem[Vignali 2003]{vi03}Vignali C., Brandt W.N., Schneider D.P., 2003, AJ, 125, 433
\bibitem[Wang 1999]{wa99}Wang T.G., Wang J.X., Brinkmann W., Matsuoka M., 1999, ApJ, 519, L35
\bibitem[Webster 1995]{we95}Webster R.L., Francis P.J., Peterson B.A., Drinkwater M.J., Masci F.J., 1995, Nature, 375, 469
\bibitem[White 1991]{wh91}White D.A., Fabian A.C., Forman W., Jones C., Stern C., 1991, ApJ, 375, 35
\bibitem[Wilkes 1987]{wi87}Wilkes B.J., Elvis M., 1987, ApJ, 323, 243
\bibitem[Wilkes 1992]{wi92}Wilkes B.J., Elvis M., Fiore F., McDowell J.C., Tananbaum H., Lawrence A., 1992, ApJ, 393, L1
\bibitem[Williams 1992]{wi92}Williams O.R. \etal, 1992, ApJ, 389, 157
\bibitem[Worrall 1987]{wo87}Worrall D.M., Giommi P., Tananbaum H., Zamorani G., 1987, ApJ, 313, 596
\bibitem[Worsley 2004a]{wo04a}Worsley M.A., Fabian A.C., Turner A.K., Celotti A., Iwasawa K., 2004a, MNRAS, 350, 207
\bibitem[Worsley 2004b]{wo04b}Worsley M.A., Fabian A.C., Celotti A., Iwasawa K., 2004b, MNRAS, 350, L67
\bibitem[Wright 199]{wr96}Wright A.E., Griffith M.R., Hunt A.J., Troup E., Burke B.F., Ekers R.D., 1996, ApJS, 103, 145
\bibitem[Yaqoob 2000]{ya00}Yaqoob T. \etal, 2000, {\it ASCA} Guest Observer Facility Calibration Memo, ASCA-CAL-00-06-01 (http://asca.gsfc.nasa.gov/docs/asca/calibration/nhparam.html)
\bibitem[Yee 1993]{ye93}Yee H.K.C., Ellingson E., 1993, ApJ, 411, 43
\bibitem[Yuan 2000]{yu04}Yuan W., Fabian A.C., Celotti A., McMahon R.G., Matsuoka M., MNRAS, 358, 432
\bibitem[Yuan 2000]{yu00}Yuan W., Matsuoka M., Wang T., Ueno S., Kubo H., Mihara T., 2000, ApJ, 545, 625 
\bibitem[Yuan 1998]{yu98}Yuan W., Brinkmann W., Siebert J., Voges W., 1998, A\&A, 330, 108
\bibitem[Zamorani 1981]{za81}Zamorani G. \etal, 1981, ApJ, 245, 357
\bibitem[Zickgraf 1997]{zi97}Zickgraf F.-J., Voges W., Krautter J., Thiering I., Appenzeller I., Mujica R., Serrano A., 1997, A\&A, 323, L21

\end{thebibliography}
\end{document}